\documentclass[usenatbib]{mnras}

\usepackage[T1]{fontenc}
\usepackage{ae,aecompl}
\usepackage{color}

\usepackage{graphicx}	
\usepackage{amsmath}	
\usepackage{amssymb}	

\newcommand{\ltsima}{$\; \buildrel < \over \sim \;$}
\newcommand{\simlt}{\lower.5ex\hbox{\ltsima}}
\newcommand{\gtsima}{$\; \buildrel > \over \sim \;$}
\newcommand{\simgt}{\lower.5ex\hbox{\gtsima}}

\title[]{The dust mass function from $z\sim$0 to $z\sim$2.5}

\author[F. Pozzi et al.]{
F. Pozzi$^{1,2},$\thanks{f.pozzi@unibo.it}F. Calura$^{2}$,G. Zamorani$^{2}$, I. Delvecchio$^{4}$,C. Gruppioni$^{2}$, P. Santini$^{3}$
\\
$^{1}$ Dipartimento di Fisica e Astronomia, Universit\`a degli Studi di Bologna, Via Berti Pichat 6/2, I--40127 Bologna, Italy \\
$^{2}$ INAF --- Osservatorio Astronomico di Bologna, Via Ranzani 1, I--40127, Italy Bologna, Italy \\
$^{3}$ INAF --- Osservatorio Astronomico di Roma, via di Frascati 33, I-00078 Monte Porzio Catone, Italy\\
$^{4}$ CEA, IRFU, DAp, AIM, Universit\`e Paris-Saclay, Universit\`e Paris Diderot, Sorbonne Paris Cit\`e, CNRS, 91191 Gif-sur-Yvette, France\\
}

\date{Accepted XXX. Received YYY; in original form ZZZ}

\pubyear{2019}

\begin{document}

\label{firstpage}
\pagerange{\pageref{firstpage}--\pageref{lastpage}}
\maketitle

\begin{abstract}
We derive for the first time the dust mass function (DMF) in a wide
redshift range, from
 $z\sim$~0.2 up to $z\sim$~2.5. In order to trace the dust emission, we start from a far-IR (160-$\mu$m)
{\it Herschel} selected catalogue in the COSMOS field. We estimate the dust masses by fitting the
far-IR data ($\lambda_{rest}$\gtsima 50$\mu$m) with a modified black
body function and we present a detailed analysis to take into account
the incompleteness in dust masses from a far-IR perspective.
By parametrizing the observed DMF with a Schechter function in the
redshift range 0.1~$<$~$z$~${\leq}$~0.25, where we are able to sample faint dust masses, we measure
a steep slope ($\alpha\sim$1.48), as found by the majority of works in the Local Universe.
We detect a strong dust mass evolution, with $M_{d}^{\star}$ at 
$z\sim$~2.5 almost one dex larger than in the local
Universe, combined with a decrease in their number density. Integrating our DMFs
we estimate the dust mass density (DMD), finding a broad peak at
$z\sim$~1, with a decrease by a 
factor of $\sim$~3 towards $z\sim$~0 and $z\sim$~2.5. In general, the trend found
for the DMD mostly agrees with the derivation of Driver et
al. (2018), another DMD determination based also on far-IR detections, 
and with other measures based on indirect tracers. 
\end{abstract}





\section{Introduction}
\label{intro}
Interstellar dust is one major component of 
galaxies, as it influences their spectral
properties across a wide range of wavelengths,
ranging from the far-infrared to the ultraviolet domain
(e.g. \citealt{1990ARA&A..28...37M}; \citealt{2009EAS....35..245D}).
A direct assessment of how the cosmic dust 
mass budget has evolved through cosmic history is of primary
importance, not only to constrain  
one major constituent of the cold mass fraction in galactic structures,
but also to have access to obscured star formation
and to the amount of heavy elements which have been subtracted
from the gas phase to end up incorporated
into solid grains (e.g. \citealt{1987ppic.proc..333T},
\citealt{1990ASPC...12..193D}, \citealt{1996ARA&A..34..279S}).
In particular, a deep understanding of the evolution of the dust mass function
(DMF) would help 
reconstructing how the buildup of interstellar dust has evolved 
in galaxies of different masses, 
possibly outlining fundamental properties, such as those shown by past 
optical studies. 
One of the most remarkable outcomes of such studies led to the discovery
of substantial differences in the star formation history of low- and high-mass galaxies (also known as 
'galaxy downsizing'), 
with a higher star formation (SF) activity in the most massive galaxies at early epochs,
followed by more intense SF in low-mass galaxies at more recent times
(e. g. \citealt{1996AJ....112..839C}, \citealt{2011MNRAS.413.2845M}). 

So far, very few studies have been dedicated to the 
evolution of the dust mass function, with direct
measurements performed only up to relatively low redshifts.   
One of the first works which have addressed this problem 
is the one of \cite{2011MNRAS.417.1510D}, in which the evolution of the space density
of galaxies as a function of their dust mass was studied up to redshift $z<0.5$. 
Their sample consisted of a sizeable amount ($\sim 2000$) of 250 $\mu$m-selected
Herschel sources from the Herschel-ATLAS Science Definition Phase, each of them 
with a reliable counterpart in the Sloan Digital Sky Survey catalogue. 
Their results indicated an increase of 
the bright end of the DMF between $z=0$ and 
$z\sim 0.5$. Such a study was complementary to another previous observational
estimate at $z\sim 2.5$, based on a much smaller and less complete sample of ultra-luminous
infrared galaxies (\citealt{2003MNRAS.341..589D}), 
broadly consistent with the DMF computed at $z=0.5$. More recently, \cite{2018MNRAS.479.1077B} determined the local
DMF by taking advantage of the
combined Herschel-ATLAS and GAMA surveys. With respect to the ones of \cite{2011MNRAS.417.1510D}, the results by \cite{2018MNRAS.479.1077B} include improved reduction of the HERSCHEL PACS and SPIRE data as well as a factor of $\sim$ 10 larger sample ($\sim$15000 sources).
The authors find an overall good match at the high-mass end of the DMF computed
by \cite{2011MNRAS.417.1510D} in the lowest redshift bin and with other estimates by other authors
(\citealt{2005MNRAS.364.1253V}, \citealt{2013MNRAS.433..695C}). \\
At the faint end, the observations of \cite{2018MNRAS.479.1077B} were much more sensitive than any previous study, allowing
to probe dust masses as low as $\sim~10^4~M_{\odot}$. In the regime of
the lowest dust masses (at $M_d<10^6~M_{\odot}$), \cite{2018MNRAS.479.1077B} derived a steeper slope of the DMF than that derived in the previous study of Dunne et al. (2011; see also \citealt{2015MNRAS.452..397C}), suggesting a larger abundance of faint dusty galaxies than expected.
More studies, based both on phenomenological arguments and observations,
have been dedicated to assessing the evolution of the comoving dust mass density.
This quantity has been estimated directly 
from the integral of the DMF by \cite{2011MNRAS.417.1510D} and \cite{2015MNRAS.452..397C},
but also by means of alternative approaches. 
For instance, \cite{2011arXiv1103.4191F} derived the total amount of dust present in the local Universe
from the integral of the cosmic star formation rate. 
\cite{2012ApJ...754..116M} estimated the cosmic density of dust residing in Mg II absorbers, visible by means
of strong absorption lines present in the spectra of distant quasars. Since strong Mg II absorbers
seem to reside mostly in galactic halos, their composition should largely trace the amount of dust which 
lies outside galaxies. 
Other authors tried to assess the comoving dust mass density of local disc galaxies (\citealt{2004ApJ...616..643F}). 
\cite{2012ApJ...760...14D} derived constraints on the comoving dust density across an extended redshift range from
unresolved sources, in particular directly from those contributing to the 
cosmic far infrared background.

More recently, \cite{2018MNRAS.475.2891D} exploited several multi-wavelength galaxy catalogues, including GAMA (\citealt{2011MNRAS.413..971D}),  G10-COSMOS (\citealt{2015MNRAS.452..616D}) and 3D-HST (\citealt{2016ApJS..225...27M}), all of them including panchromatic photometric
information ranging from the UV to the mid-IR. 
By means of the MAGPHYS code (\citealt{2008MNRAS.388.1595D}) and with
computationally intensive SED-fitting techniques, these authors attempted to model the spectral energy distribution of 
galaxies at various redshifts, based on the energetic balance between the radiation attenuated by dust
in the UV and optical bands and the amount re-radiated in the far-IR.\\
On the theoretical side, these results have been interpreted by means of
cosmological models (e. g. \citealt{2017MNRAS.471.3152P}) and 
non-cosmological theoretical approaches (e. g. \citealt{2017MNRAS.471.4615G}), 
but the global picture emerging from such theoretical studies is far from being clear. 
For instance, the bulk of the comoving dust mass estimated in the local Universe seems to lie outside galaxies
(\citealt{2017MNRAS.471.4615G} and references therein), 
but cosmological simulations indicate that locally the amount of grains contained in galactic structures should be dominant,
strengthening the idea that intergalactic dust can hardly survive against destruction mechanisms such as sputtering (\citealt{2018MNRAS.478.4905A}).  

In order to improve our current understanding of the origin of cosmic dust,  
an updated observational assessment of the amount of dust present in galaxies and calculated across an extended redshift range
is required, and represents the aim of the present work. 

We exploit the Herschel catalogue to perform a direct, possibly unbiased measurement
of the evolution of the dust mass function up to redshift 2.5.
By extending the study of \cite{2011MNRAS.417.1510D} across a large redshift range, 
we hope to gain a clearer view of the contribution of resolved sources to the cosmic dust budget, covering a significant fraction of cosmic time. The size and wealth of multi-wavelenth information of galaxies from 
the PEP Herschel survey (\cite{2011A&A...532A..90L} lying in the 
COSMOS field render the sample considered in this study an ideal tool to perform such a task. 

This paper is organized as follows. In Section 2, we present the main features of the dataset and the main assumptions 
used to derive our estimates of the dust mass in galaxies. In Section
3 we present our results, i.e. our dust mass functions and dust mass density. Finally, in Sect. 4 we draw our conclusions.

\section{Sample Description}
\label{sample_sec}
The catalogue used in this work is based on a far-IR sample 
selected in the  ${\sim}$2 deg$^{2}$-wide COSMOS field and obtained
within the {\it Herschel}-PEP survey
(\citealt{2011A&A...532A..90L}). We consider the latest released blind catalogue selected
at 160-${\mu}$m (DR1, 7047 sources) with $>3{\sigma}$ flux density,
corresponding to a flux limit of ${\sim}$9.8 mJy.
The choice of considering as parent sample a far-IR catalogue is
guided by the necessity of having several detections at different
wavelengths for each system to constrain
the dust masses, and yet with a very simple selection function  (see Sec. \ref{method_sec}).
From the original 160-${\mu}$m selection, we built a multi-band catalogue taking advantage of the extensive
multi-wavelength coverage in the COSMOS field. Concerning the other far-IR PACS band (100 $\mu$m) and the mid-IR 24-${\mu}$m band, we use the association available in the DR1 release and based on the maximuum
likelihood technique (\citealp{1992MNRAS.259..413S},
\citealp{2001Ap&SS.276..957C}). For the cross-match with the SPIRE
far-IR bands (250, 350, and 500 ${\mu}$m) we used the same catalogue considered in previous PACS-based works
(i.e. \citealt{2013MNRAS.432...23G}, \citealt{2015MNRAS.449..373D})),
the ones provided by the HerMES collaboration (\citealt{2010MNRAS.409...48R}) using the {\it Spitzer}-MIPS 24-$\mu$m
positions as priors to extract the SPIRE fluxes. Finally, the IRAC/optical/UV fluxes taken from the
COSMOS2015 catalogue (\citealp{2016ApJS..224...24L}) were merged to
the PACS-160-${\mu}$m sample by matching the 24-$\mu$m counterparts listed in both. The COSMOS2015 is NIR selected, where
objects have been detected from the sum of the UltraVISTA-DR2 $YJHK$ and $z^{++}$ images. By
construction, in comparison to the previous $i$-selected catalogue,
this catalogue is missing a
fraction of blue, faint, star-forming galaxies
(\citealt{2016ApJS..224...24L}). For this reason, we decided to
cross-match the far-IR sources with no counterparts in the
COSMOS2015 catalogue with the \cite{2009ApJ...690.1236I} $i$-selected
catalogue. Totally, among the 160-${\mu}$m selected sources (7047),
6002 are with 24-$\mu$m counterparts (${\sim}86\%$), of which 5993 with
available NIR or optical counterparts (${\sim}99.9\%$, 5783 in the COSMOS2015 and 210
in the \citealt{2009ApJ...690.1236I} catalogue). While the 
cross-matching with the optical/NIR bands does not involve almost any
source loss, a moderate (14$\%$) but
not negligible fraction of the far-IR sources does not have 24-$\mu$m
counterparts. A fraction of these sources are likely spurious
sources, as shown by the simulations done for the DR1
release PEP catalogue ($\sim$5\% at the 3$\sigma$ flux
level). In Sec. \ref{method_sec} we will describe
our method to correct for incompleteness and for the presence of spurious systems.

We assigned a redshift measurement to each source, either
spectroscopic ($\sim50\%$ of the sample) or photometric. For the COSMOS2015 (\citealt{2016ApJS..224...24L}) 
and the $i$-band (\citealt{2009ApJ...690.1236I}) counterparts, the
photometric redshifts were taken from their respective catalogues.
These were computed via the Le Phare SED-fitting code
(\citealt{2002MNRAS.329..355A}; \citealt{2006A&A...457..841I}, as
described in \citealt{2013A&A...556A..55I}). As for the
photometric redshifts taken from the COSMOS2015 catalogue
(the vast majority of our sources) at the median magnitudes of our
sample ($i\sim$21.6), we benefit of a very high accuracy
(${\sigma}_{\Delta{z}/(1+z)}\sim{0.007}$, catastrophic failure
$\eta$=0.5 $\%$, see Table 5 from \citet{2016ApJS..224...24L}).

For X-ray detected sources (352 objects), we used a separate set of photometric
redshifts from \cite{2016ApJ...817...34M}
which were derived via SED fitting, with templates which include galaxies, AGN/galaxy hybrids, AGNs, and QSOs.
When available, we used spectroscopic redshift measurements taken from an exhaustive list
made internally acccessible to the COSMOS team (Salvato et al., in prep.).


\begin{figure*}
\includegraphics[width=0.7\textwidth,angle=270]{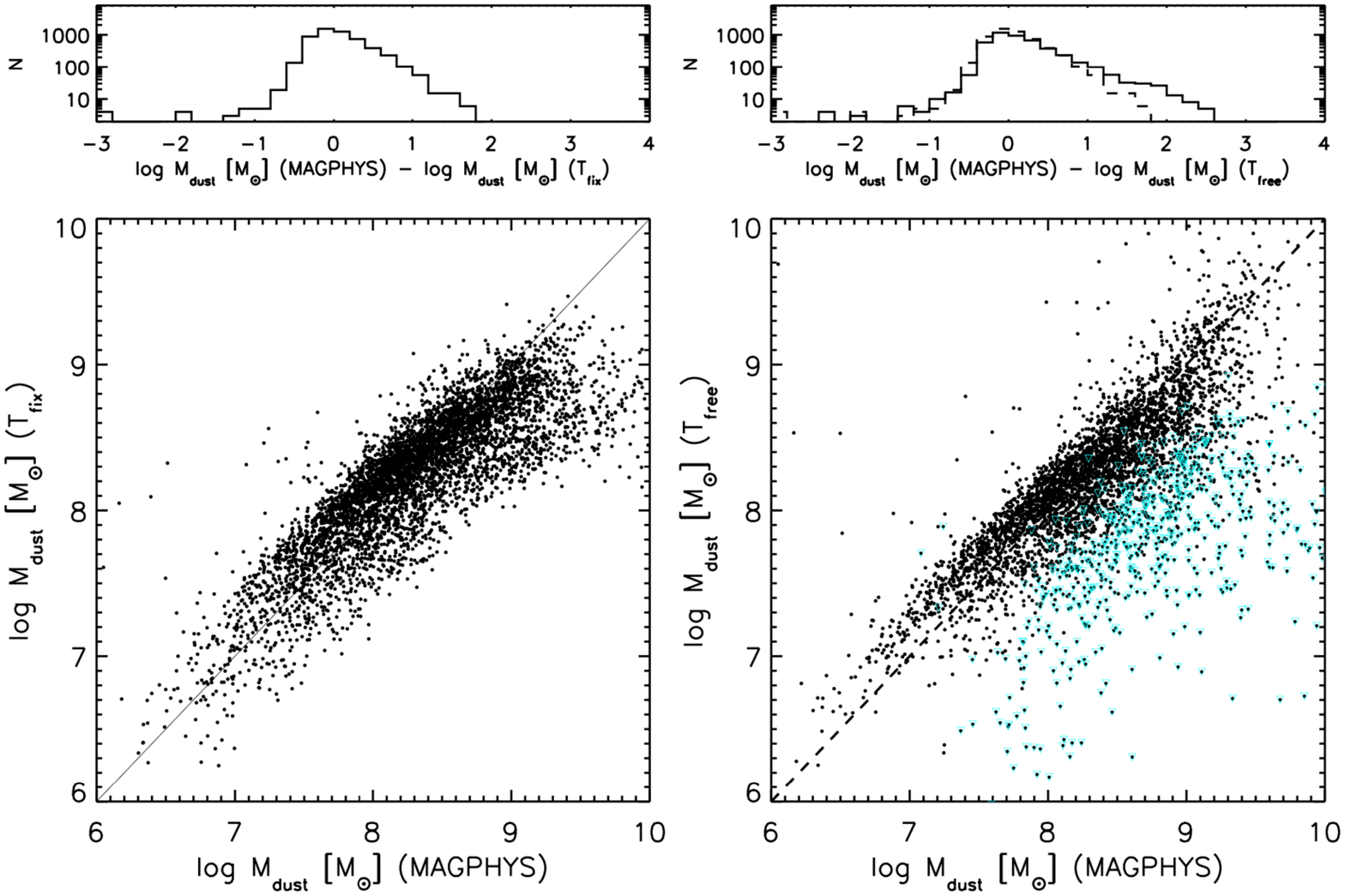}
\caption{(Left panel) Dust masses derived by means of the MBB fit by fixing the temperature according to the relation found
by Magnelli et al. (2014) versus the dust masses derived with the MAGPHYS cod e (da Cunha et al. 2008). The above 
histogram shows the difference between the dust mass estimates.  
The value of the dispersion  of the distribution is  $\sigma=0.4$. 
(Right panel) Dust masses derived by means of  the MBB fit and by leaving the temperature as a free
parameter versus the dust masses derived with the MAGPHYS code. 
The cyan points represent galaxies with fitted temperature values $T>40$ K. 
As for the right figure, above histogram 
shows the difference between the
two dust mass estimates (solid line) compared 
woth the difference obtained in the left panel (dashed line). 
The histogram obtained by leaving the temperature as a free parameter has
a tail which extends rightwards by $\sim$2.5 dex. 
In both figures, the lines represent the 1:1 relations.}
\label{figure_mdust}
\end{figure*}

\begin{figure}
\includegraphics[width=0.35\textwidth,angle=270]{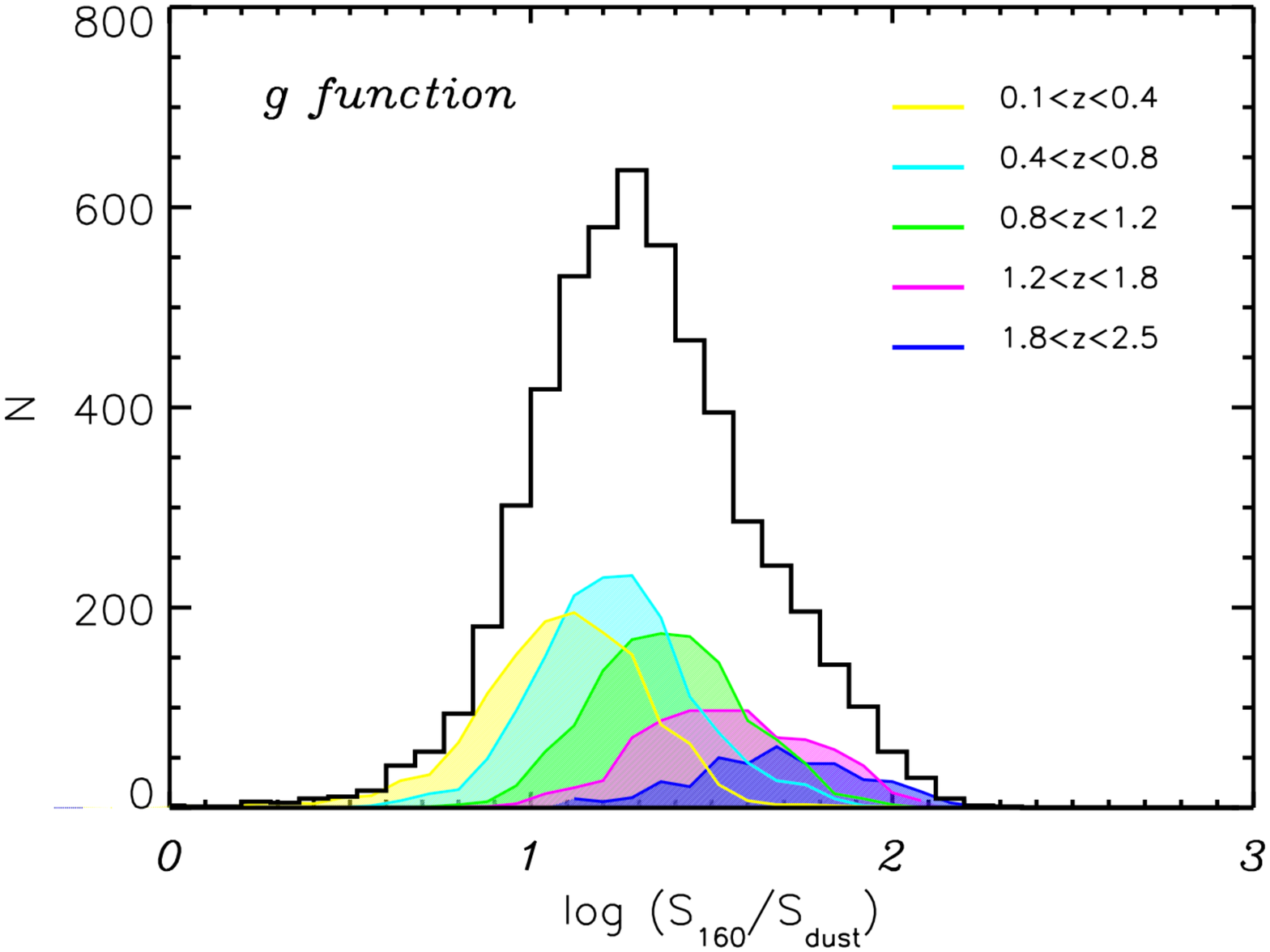}
\caption{Distribution of the $log(S_{160}/S_{dust})$ ratio (called
  'g-distribution' in the text) for all the
  galaxies (black line) and splitting the sample in different redshift
  bins (different colors as indicated in the legend). }
\label{figure_gfunc}
\end{figure}

\subsection{Dust masses and temperatures}
\label{dust_masses_sec}

In order to estimate the dust mass density we need an accurate measure of
the dust masses for the galaxies, giving particular
attention to the possible presence of systematic selection biases.\\
In the last years, mainly thanks to the {\it Herschel} satellite, many
physically-motivated models have been
developed for deriving fundamental galaxy quantities (such as the stellar
mass, the star-formation rate and the mass and temperature of
the dust) from SED-fitting to photometric data. Among these are the CIGALE (\citealt{2005MNRAS.360.1413B},\citealt{2019A&A...622A.103B}) and MAGPHYS
codes (\citealt{2008MNRAS.388.1595D}), based on energy
balance equilibrium, and the GRASIL code (\citealt{1998ApJ...509..103S}),
the only model that takes into account also the relative 
spatial distribution of stars and dust through a radiative transfer
code. These codes model the
complexity of the dust in the ISM (different temperatures,
geometry, composition and radiation
fields) but can suffer by parameters' degeneracies when the
observed spectro-photometric information is not exhaustive (see
discussion in \citealt{2018A&A...609A..30S} and \citealt{2019arXiv190204851H}). A different and simpler approach for
deriving the dust integrated properties (temperature and mass) comes
from assuming that the bulk of the dust in galaxies is heated at thermal
equilibrium by a mean interstellar radiation field (\citealt{2013A&A...552A..89B}).

Given its simplicity, a single temperature dust component is the approach we have chosen for
our analysis by fitting the data points with a modified black-body (MBB)
spectrum and deriving the dust masses using the relation
valid in the standard optically thin regime (i.e. \citealt{2013A&A...552A..89B}),

\begin{equation}
\label{eq_mdust}
M_{d}=\frac{D_{L}^{2}S_{\nu_{obs}}}{ (1+z)k_{\nu}B_{\nu}(T)}
\end{equation}

\noindent where ${\nu}$ and ${\nu_{obs}}$ are the rest-frame and
observed frequency ($\nu={\nu}_{obs}(1+z)$), $B_{\nu}(T)$ is the Planck function, $D_{L}$ the luminosity distance, $S_{\nu_{obs}}$ is the observed flux
corresponding to a rest-frame frequency of 1.2 THz (250 $\mu$m). Following \cite{2013A&A...552A..89B}, we have adopted
$k_{\nu}=4.0(\frac{\nu}{1.2THz})^{\beta}$ cm$^2$gr$^{-1}$ and $\beta=2$ (see also \citealt{2014A&A...562A..67G}).

Eq.~\ref{eq_mdust} has 2 free parameters, the temperature
$T$ of the dust and the normalization, directly linked to the dust mass
$M_{d}$. Consequently, we apply this method to galaxies
with at least 3 photometric points in the far-IR regime ($\lambda_{rest}>50$
$\mu$m). 

The adopted method accounts only for the cold diffuse dust, while
we know that the dust resides in galaxies at different temperatures
(i.e. the cold diffuse component, the warm component around birth
clouds and the hot dust transiently heated to temperatures $> 50 ~K$). 
However, the diffuse cold
component accounts for the bulk of the dust mass (\citealt{2011MNRAS.417.1510D}) and, consequently,
the adopted procedure gives robust dust mass estimates. This is shown, for example,
in \cite{2013A&A...552A..89B}, where a detailed comparison of the dust masses
derived by different approaches has been presented taking advantage of the full sampling
of the dust emission in local galaxies from the KINGFISH
survey (\citealt{2011PASP..123.1347K}): a MBB fit can give
 dust masses consistent with those derived by using a full spectral
 energy distribution of the dust emission (\citealt{2007ApJ...657..810D}). 
In \cite{2011MNRAS.417.1510D}, a second warm component is
added to the cold one, but this component is found to account only for
10 \% of the total dust budget. 

The reliability of the MBB approach has been shown also recently by \cite{2019arXiv190204851H},
where a detailed comparison of the physical galaxy parameters
derived by different codes is reported. The MBB approximation gives dust masses
consistent with those derived by physically motivated-models with
a scatter $\sigma$\ltsima{0.15} dex.

A factor that can greatly influence the dust masses is linked to the
adopted selection procedure. 
The cold dust is typically at $\sim$25-40 K (\citealt{2014A&A...561A..86M}) and its emission peaks at around
100-120 $\mu$m. This means that moving in redshift, different rest-frame parts of the
dust spectrum are sampled with the 160-$\mu$m data, i. e. the Rayleigh-Jeans region at
$z${\ltsima}~0.5, the dust peak at $z\sim$0.5 and the Wien
region at $z$\gtsima1. This implies that at $z$\gtsima 1 the measure of the
dust temperature can be overestimated and, as the dust mass is strictly
dependent on the temperature via 
Eq. \ref{eq_mdust}, this can introduce a
bias in the estimated dust mass. To overcome this, we have fixed the temperatures of our
galaxies to
the values expected from the relation found by
\cite{2014A&A...561A..86M} for star-forming galaxies on the basis of
their redshift and their specific star-formation rate (SFR/M$_{\star}$) 
($T_{dust}=98{\times}(1+z)^{-0.065}+6.9{\times}log(SFR/M_{\star}$). Taking
into account the evolution of the specific star-formation rate with
redshift (i.e. \citealt{2014ApJS..214...15S}),
\cite{2014A&A...561A..86M} found a net positive evolution of the
temperature as a function of the redshift (see also
\citealt{2015A&A...573A.113B} and \citealt{2018A&A...609A..30S}), and the
relation they obtain is less contaminated than other results by
selection biases since they performed their analysis using both individual and stacked {\it Herschel} images.


Before proceeding, we underline that although a far-IR selection can be subject to critical aspects concerning
the dust temperature measures, it is however the best band for tracing the IR
luminosity of star-forming galaxies, both in terms of completeness of
the sample and accuracy of the IR luminosity estimates (see \citealt{2019MNRAS.483.1993G}).

Once the temperatures from the \cite{2014A&A...561A..86M} relation are assigned to
our galaxies, we derive from the SED-fitting procedure previously described the
normalization of the MBB function and from Eq. \ref{eq_mdust} the
dust masses. 

As a further check of the  accuracy of the adopted method, in
Fig. \ref{figure_mdust} (Left Panel) we compare our fiducial values of
the dust masses (i.e. derived by fixing the temperature) with the dust
masses derived using a completely different approach, i.e. using the MAGPHYS
code (\citealt{2008MNRAS.388.1595D}). This code performs an
SED-fitting considering all the photometric data (from UV to far-IR) and is based on the energy
balance between the energy absorbed in the UV/optical band due to dust 
and re-emitted in the mid- and far-IR. We find that the dust
masses derived with the adopted method are in general agreement
($\sigma\sim0.4$ dex with respect to the 1:1 relation, see upper histogram) with the MAGPHYS
estimates.
However, a bending at high dust masses is present in our distribution, with our
fiducial values showing a plateau at ${\sim}10^{9}$ M$_{\odot}$, 
which is not present in the estimates derived with the MAGPHYS
code.

Differences are expected between estimates derived with 
different methods. For instance, here we consider only one dust 
component, whose temperature, using the relation from \cite{2014A&A...561A..86M},
is in the range 25-35 K; in MAGPHYS, on the contrary, emission from
three dust components (hot, warm and cold) is considered. The cold
component can have a temperature as low as 15 K and the dust
mass is inversely proportional to the temperature: this can be one reason
for the higher masses.

From a more theoretical point of
view, chemical evolutionary models have difficulties
in reproducing dust masses higher than ${\sim}10^{9}$
M$_{\odot}$, even considering extreme physical conditions (i.e. a top-heavy
Initial Mass Function (IMF), see \citealt{2017MNRAS.465...54C}). 

In Fig. \ref{figure_mdust} (Right Panel), the same comparison is 
shown for the dust masses derived by leaving the temperatures as a
free parameter in our fitting. The number of outliers increases (i.e. the percentage
of galaxies with dust masses different by more than
one order of magnitude goes from $\sim4\%$
to $\sim{8}\%$, ) and galaxies with $T>40$ K (as shown by the cyan points) have 
systematic lower masses up to a factor of ten or more. As previously anticipated,
this is likely due to unphysically high temperatures due to selection
biases. The increase of the number of outliers is evident in the upper
histogram, where the differences between the masses
estimated with the MAGPHYS code and the masses estimated in
case of leaving the temperature as a free parameter (solid line) and
by fixing it (dashed-line) are reported.

In future studies, it will be important to improve our estimate of the
effects of the temperature on the derived dust mass. For instance,
one could explore the use of a prior
distribution, i.e., a Gaussian function centred on the empirical values,
also allowing
for some dispersion and considering their effects on the derived values.

For a very low fraction of the sources (6 \%), there are not 3 photometric far-IR points or the adopted fitting
procedure fails to find a stable solution. We account for these
sources in the completeness correction (see Sec. \ref{method_sec}).

\begin{figure}
\centering
\includegraphics[width=0.5\textwidth]{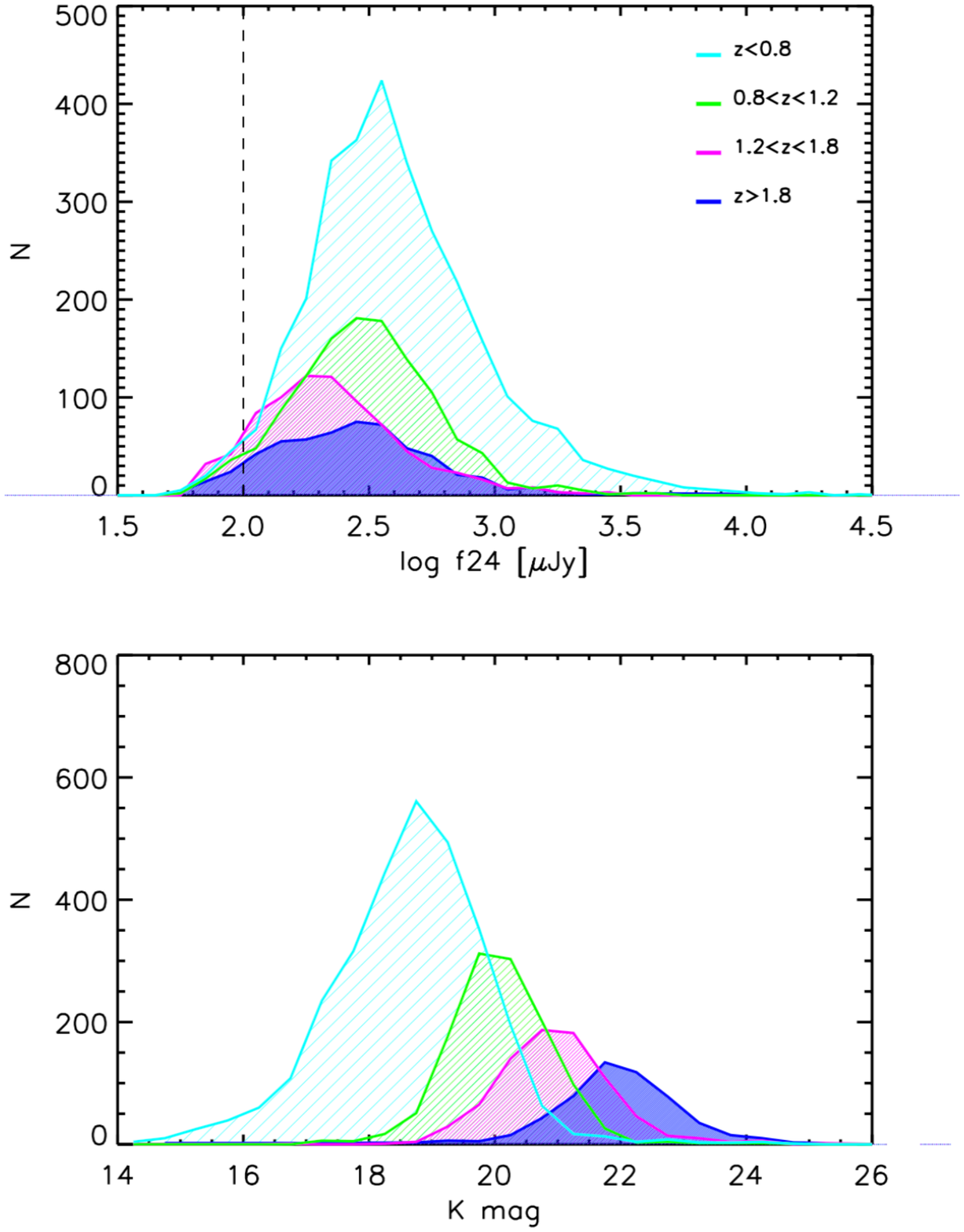}
\caption{(Top panel) 24-$\mu$m flux distribution of the 160-$\mu$m targets with an associated 24-$\mu$m counterparts. The
  distribution of the sample divided
  in four redshift bins ($z<$0.8, 0.8$<z<$1.2, 1.2$<z<$1.8 and z$>$1.8) are shown as colour filled regions
. (Bottom panel). The dashed line indicates the maximum 24-$\mu$m flux
value we have considered to correct for the incompleteness due to the
presence of 160-$\mu$m sources with no 24-$\mu$m counterparts (see Sec. 3.1 for details). (Bottom panel): $K$-band
magnitude distribution of the NIR counterparts of the 160-$\mu$m
targets. The $K$-band magnitudes are from the COSMOS2015 (Laigle et al 2016) and
the Ilbert et al. (2009) catalogues. The lines and the colour filled
regions are as in the Top panel.}
\label{figure_comple}
\end{figure}



\section{Dust mass function}

In this section we investigate the evolution of the dust mass
function with redshift. The {\it Herschel}
sources for which we where able to recover $M_{d}$ are 5546 (92\% of
the far-IR sample with 24-$\mu$m counterparts). 

\subsection{Method Description}
\label{method_sec}

We use the non-parametric $1/V_{max}$ method
(\citealt{1968ApJ...151..393S}) for the computation of the dust mass function.
This method is widely used in literature when a physical parameter is derived directly from 
the data, as the luminosity from flux densities. Following 
\cite{2004A&A...424...23F} (see also \citealt{2014MNRAS.439.2736D}) the $1/V_{max}$ method
can be used also when the number density of an indirect physical
parameter (as the dust mass in the present case) is investigated starting
from a flux density selection (the 160-$\mu$m band).

In the `standard' case of a luminosity function, for each source of
the sample $V_{max}$ is the volume corresponding to the the largest distance at which that source would be detected and 
is estimated as 

\begin{equation}
V_{max}=\int_{z_{min}}^{z_{max}}{\frac{dV}{dz}\Omega(z)dz}
\end{equation}

\noindent with $z_{min}$ being the lower boundary of a given redshift
bin and $z_{max}$ the minimum value between the upper boundary of the
redshift bin and the maximum redshift at which the source would be
detected.  $\Omega(z)$ is the effective area of visibility of the
source and is defined as

\begin{equation}
\Omega(z)=\Omega_{geom}f_{c}(z)
\end{equation}

\noindent where $\Omega_{geom}$ is the geometrical projected sky
area observed and $f_{c}(z)$ the completeness function at a given
redshift. Given the $V_{max}$ value estimated for each source, the
luminosity function  is
computed as 
\begin{equation}
\Phi(L,z)=\frac{1} {{\Delta}logL}{\sum_{i=1}^{n}\frac{1}{V_{max,i}}}
\end{equation}

\noindent  where $L$ is the luminosity and ${\Delta}logL$ is the size of the luminosity bin. 
When the number density of an indirect physical parameter, such as the
dust masses in our case, needs to 
be estimated, the $1/V_{max}$ method can still be applied (replacing
the mass to the luminosity) once a
proper evaluation of the incompleteness function $f_{c}(z)$ is carried
out. We define the quantity
$S_{dust}=\frac{M_{dust}}{4{\pi}D_{L}^{2}}$ and we compare $S_{dust}$ with
the corresponding 160-$\mu$m flux to measure how our selection at
160-$\mu$m affects the selection in mass. The same procedure was
used by \cite{2014MNRAS.439.2736D} (see their Appendix 
A), where the accretion power of AGNs was evaluated from a sample of
far-IR selected galaxies. 

In Fig. \ref{figure_gfunc} the distribution $g$ of the flux ratios
$S_{160}/S_{dust}$ is plotted for all the galaxies of our sample
(black line) and dividing the galaxies according to their
redshifts. As clearly shown in the figure, the distribution shifts towards higher values with increasing redshift.
Assuming that the $g$ distributions shown in Fig. \ref{figure_gfunc}
keep the same shape even at fluxes lower than the $S_{160}$ detection
limit, we estimated the completeness fraction ($f_c$) at
160-$\mu$m as a function of $S_{dust}$. Specifically, $f_c$ was
calculated by convolving the $S_{160}$ source counts with the observed
distribution $g$ of the flux ratios $S_{160}/S_{dust}$ as follows:

\begin{equation}
\label{eq_fdet}
f_c(logS_{dust})=\frac{\int_{logS160,lim}^{logS160,max}{\frac{dN_{exp}}{dlogS_{160}}g(x)dlogS_{160}}}{\int_{logS160,min}^{logS160,max}{\frac{dN_{exp}}{dlogS_{160}}g(x)dlogS_{160}}}
\end{equation}

\noindent where $\frac{dN_{exp}}{dlogS_{160}}$ represents the 160-$\mu$m
source counts and $g(x)$ is the $S_{160}/S_{dust}$ distribution
interpolated at the flux ratio corresponding to
$x=log(S_{160}/S_{dust})$ (see Fig. \ref{figure_gfunc}). The source
counts at 160-$\mu$m in Eq. \ref{eq_fdet} are the observed (numerator)
and the intrinsic (denominator) source counts depending on the lower
end of integration: for the observed counts, we set the minimum of the 160-$\mu$m flux at the 160-$\mu$m
detection limit, while for the intrinsic counts we stop the
integration when the
derived $f_c$ is equal to 0.5. This latter value coincides with
160-$\mu$m fluxes at which the number of detections and non-detections are the
same at a given $S_{dust}$. 
Before estimating the dust mass density, a further incompleteness
correction must be applied since $\sim$14\% of the 160-$\mu$m sources do not
have a 24-$\mu$m or optical counterpart. Since likely $\sim$5\% of them
are spurious, as stated in Sec. \ref{sample_sec}, the `real' 160-$\mu$m
sources with no optical counterpart we must consider and distribute
among the redshift bins are $\sim$9\% of the sample. In
Fig. \ref{figure_comple} the 24-$\mu$m flux distribution of the
160-$\mu$m targets with 24-$\mu$m counterpart are shown, divided in
four redshift bins. 
As clearly shown from the figure, the sources
with a faint 24-$\mu$m counterpart, near the 24-$\mu$m limit, are not
all at high-$z$,
but a tail of faint sources is present also at lower redshift. Considering 24-$\mu$m flux values \ltsima
200 $\mu$Jy, the sources at 0 $<z<$ 0.8, 0.8 $<z<$ 1.2,  1.2 $<z<$ 1.8
and $z>$1.8 are 35 \%, 22\%, 22\% and 17\%, respectively. 
These fractions have also been used to re-distribute
all the 160-$\mu$m sources with no 24-$\mu$m counterpart 
in the same four redshift bins. 
This corresponds to the multiplication of the DMF at 0 $<z<$ 0.8, 0.8 $<z<$ 1.2,  1.2 $<z<$ 1.8
and $z>$1.8 by a factor 1.08, 1.13, 1.23 and 1.21, respectively. In
Fig. \ref{figure_comple} we show also the K-band distribution of the
  160-$\mu$m sources with 24-$\mu$m and a near-IR
  identification (i.e. the sources for which having at least a value for the photometric redshift we are able to recover
  the dust masses). The faintest 
  sources in the COSMOS2015 catalogues are at K$\sim$25-26, while most
  of the
  K-band counterparts of our sample are almost two magnitudes
  brighter. This guarantees that almost all
  the sources have a near-IR counterparts, as indicated by the very
  high percentage of identifications (nearly 100\%, see Sec. \ref{sample_sec}).
Finally, in order to take into account the few failures of the
fitting procedure and the galaxies with less then 3 photometric far-IR
points (see Sec. \ref{dust_masses_sec}), we apply a uniform
correction (further multiplying by 1.05 the DMF in each redshift bin). 


\begin{figure}
\centering
\includegraphics[width=0.35\textwidth,angle=270]{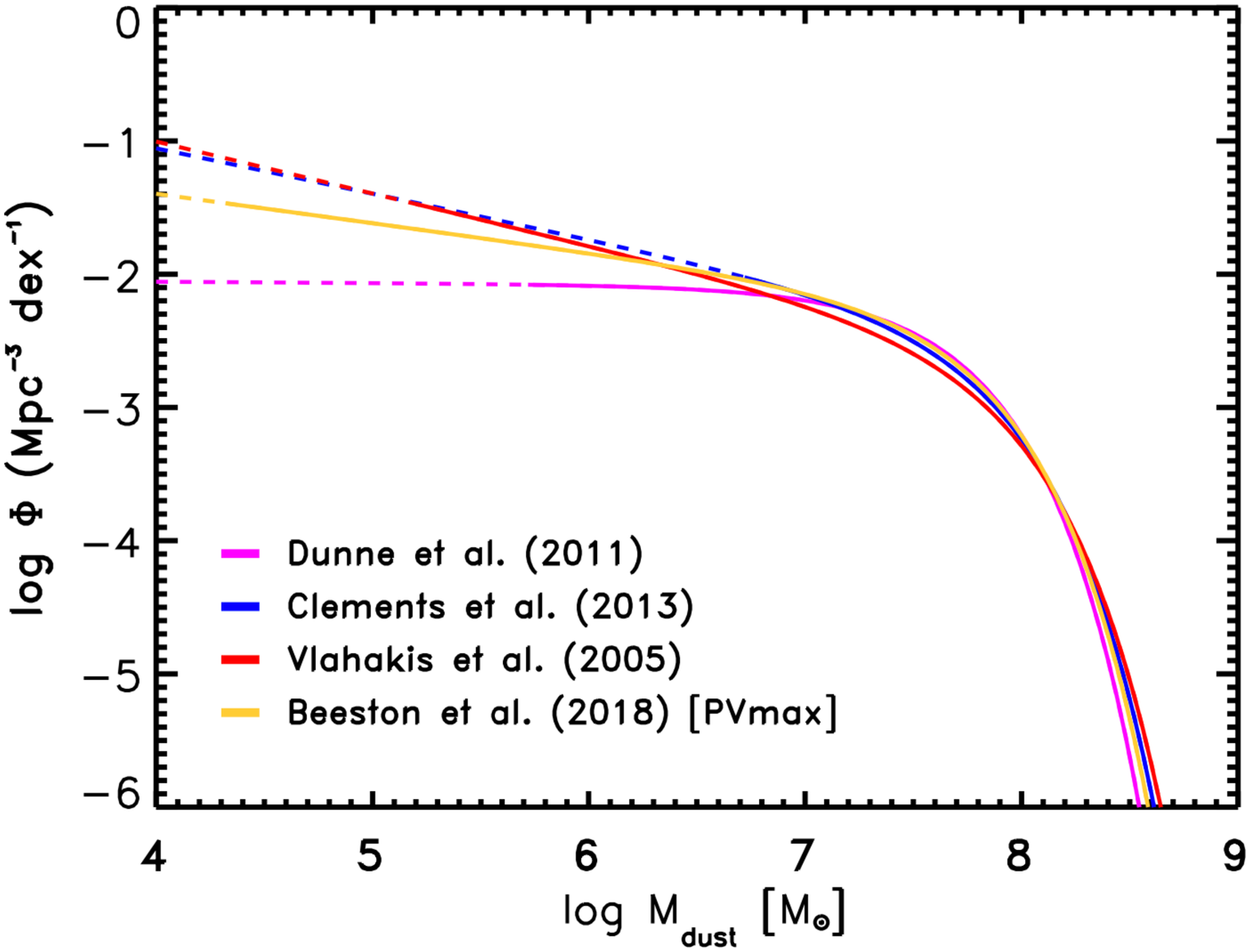}
\caption{Local DMFs from the literature. The Schechter fit
  parameters are taken from Table 1 of Beeston et al. (2018).
  The following determinations are reported: Clemens et al. (2013) from $\sim$200 galaxies from the
  all-sky bright Planck catalogue (blue line); Vlahakis et
  al. (2005) from ground-based submm
  measurements (SCUBA) of local optically selected galaxies (red line);
  Dunne et al. (2011), from $\sim$2000 250-$\mu$m
  selected galaxies from the Science Demonstration Phase of
  the Herschel-ATLAS (magenta line) and finally of Beeston et al. (2018) from
  a large sample ($\sim$15000) galaxies from the GAMA and H-ATLAS
  surveys (orange line). The Dunne et al. (2011) DMF includes the correction factor of
  1.42 for the density of the GAMA09 field (see Beeston et
  al. 2018). Solid (dashed) lines represent the Schechter functions
  measured (extrapolated) for each determinations. }
\label{figure_dmf_local}
\end{figure}

\begin{figure*}
\centering
\includegraphics[width=0.7\textwidth,angle=270]{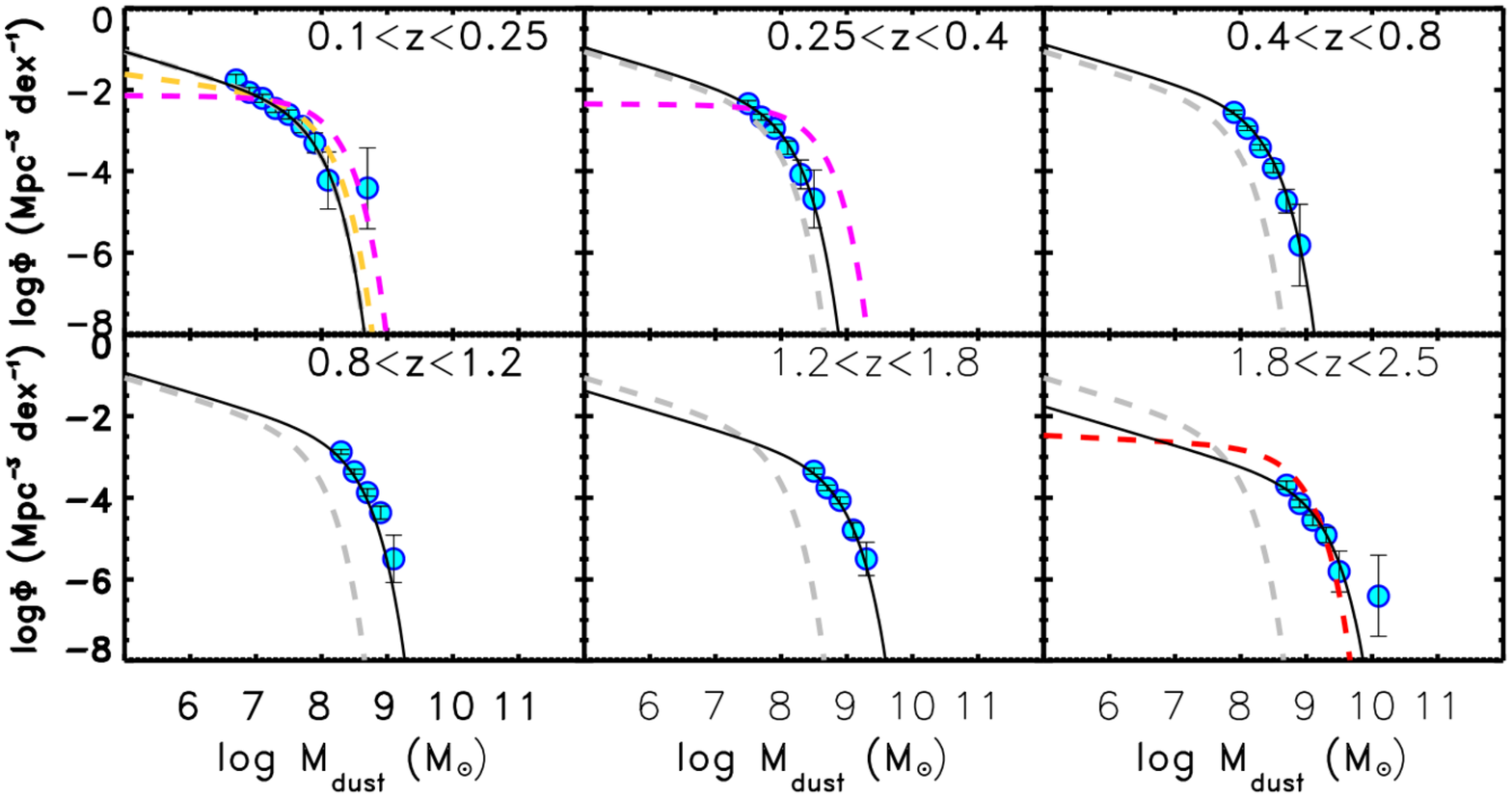}
\caption{DMFs in six redshift bins estimated using the non-parametric
  $1/V_{max}$ method (cyan points). The errors represent the
  $\pm1\sigma$ (Poissonian) uncertainties. The best-fitted Schechter
  functions are overplotted to the data points with the faint-end
  slope fixed to that which fits best in the 0.1$<z<$0.25 redshift
  bin. Parameters for the fits are given in Table \ref{table1}. In
  each panel the best-fit DMF obtained in the first bin is reported
  for comparison as a dashed grey line. The
  dashed magenta line in the first and second panels represent the DMF from Dunne
  et al. (2011) in the redshift bins 0.1$<z<$0.2 and 0.2$<z<$0.3, respectively, from the
  Herschel-ATLAS Science Definition Phase; the yellow line is the DMF from Beeston et al. (2018) at $z<$0.1 from the
    combined Herschel-ATLAS and GAMA surveys. The Dunne
  et al. (2011) DMF includes the correction factor of
  1.42 for the density of the GAMA09 field (see Beeston et al. 2018). The
  dashed red line in the last panel represents the DMF from Dunne
  et al. (2003) in the redshift bin 1$<z<$5. }
\label{figure_vmax}
\end{figure*}


\subsection{Dust mass function best-fit}
\label{dmf_fit_sec}

In the local Universe, many estimates of the dust mass functions
 have been derived using different methods (see Fig. \ref{figure_dmf_local}); our work, for the first
time, extends the measure of the DMF at high-$z$, $z{\sim}2.5$. In the
literature, the only other determinations of the DMF beyond the local Universe are from
\cite{2011MNRAS.417.1510D} up to $z\sim{0.5}$ and their previous
estimate at $z\sim{2.5}$ (\citealt{2003MNRAS.341..589D}). The DMF from
\cite{2011MNRAS.417.1510D} was obtained using
$\sim{2000}$ galaxies selected at 250-$\mu$m from the Herschel-ATLAS survey and their results indicate an increase of the bright end of the
DMF between $z\sim{0}$ and $z\sim{0.5}$, with the most massive galaxies at
$z\sim{0.5}$ having dust masses about a factor of five larger than
those at $z\sim{0}$. 

In Fig. \ref{figure_vmax} we report our estimates of the dust mass function (DMF) in 6 redshift bins: 0.1~$<$~$z$~${\leq}$~0.25, 0.25~$<$~$z$~${\leq}$~0.4,
0.4~$<$~$z$~${\leq}$~0.8, 0.8~$<$~$z$~${\leq}$~1.2,
1.2~$<$~$z$~${\leq}$~1.8 and 1.8~$<$~$z$~${\leq}$~2.5. The
width of the redshift bins was chosen in order to include a large number (\gtsima
800, except the first and last bins with $\sim$535 and 435 objects, see Table \ref{table1})
of sources in each of them. 
The reported errors represent the ${\pm}1\sigma$
Poisson uncertainties. We have chosen not to explore the DMF at lower (higher)
redshift because of the small number of sources 
($\sim$100 both at $z$~$<$~0.1 and $z$~$>$~2.5) and because of 
the large and uncertain incompleteness corrections, especially at $z$~$>$~2.5. 

The first panel of Fig. \ref{figure_vmax} shows that our best
fit of the DMF (solid black line) computed at the median redshift of
the bin, $z=0.15$, is only marginally consistent with the one of
\cite{2011MNRAS.417.1510D} in their redshift bin ($0.1<z<0.2$) (pink dashed line).
The \cite{2011MNRAS.417.1510D} DMF shows both a higher bright-end by up to 
a factor of 10 at M$_{d}$\gtsima${10^{8.5}}$ M$_{\odot}$ and a flatter
faint end slope. It is difficult to address 
the cause of the differences; however, we note that 
\cite{2011MNRAS.417.1510D} associate the optical
counterparts directly to the 250-$\mu$m parent
catalogue, while we go through an intermediate step, associating our
160-$\mu$m selected sources to 24-$\mu$m sources. Our procedure is likely
to enhance the probability of associating the right counterparts to our systems, 
since we are taking into account not only the
relative distances between the FIR sources and the counterparts, but also
the typical SED of a star-forming galaxy. Moreover, the difference in the faint-end slope is
only indicative, since in \cite{2011MNRAS.417.1510D} the slope is not
directly observed at $z\sim{0.15}$ but is
extrapolated from their determination at lower ($z\sim{0.05}$) redshift (at $z\sim{0.15}$ the
DMF of \citealt{2011MNRAS.417.1510D} probes only dust
masses larger than M$_{d}$\gtsima${10^{7.3}}$ M$_{\odot}$). In the same panel we also report the local determination from
\cite{2018MNRAS.479.1077B} (yellow dashed line) which, as said before, is an updated
determination of the DMF from the Herschel-Atlas survey.
 The \cite{2018MNRAS.479.1077B} DMF is consistent with
 ours at the bright-end and has a faint-end slope steeper than
 \citealt{2011MNRAS.417.1510D} ($\alpha{\sim}$1.3 instead of
 $\alpha\sim$1.0, see \citealt{2018MNRAS.479.1077B}) more consistent
 with our data. In the second plot, our DMF computed at the median redshift of
$z=0.32$ is compared with the \cite{2011MNRAS.417.1510D} in their redshift bin ($0.2<z<0.3$) (pink
dashed line). The same differences observed at $z=0.15$ are evident, with
the \cite{2011MNRAS.417.1510D} determination having a larger bright-end value and a
flatter faint-end slope.

The DMF from \cite{2003MNRAS.341..589D} at $z\sim{2.5}$ (overplotted
as red dashed line to our data in the last panel of Fig \ref{figure_vmax}) had been
estimated from sub-mm SCUBA data, before the advent of
the Herschel IR mission. Despite the different underlying assumptions,
such as taking a unique SED for all galaxies (i. e. the one of Arp220) and the
large redshift bin adopted ($1<z<5$), the two
determinations are in good agreement.

In order to parametrize the DMF and estimate its evolution, we fit the
$1/V_{max}$ data 
points with a Schechter function
(\citealt{1976ApJ...203..297S}) of the form:

\begin{equation}
\Phi(M_d)dlogM_d=\Phi_d^{\star}e^{-{\frac{M_d}{M_{d}^{\star}}}}{\frac{M_d}{M_{d}^{\star}}^{-\alpha+1}}dlogM_d
\end{equation}

\noindent where the best-fit values for the three free parameters ($\alpha$, $M_{d\star}$ and $\Phi_d^{\star}$) are derived
by means of a non linear least square fitting procedure. 
The definition of  $\Phi_d^{\star}$ (\citealt{2018MNRAS.479.1077B}) incorporates the factor {\it ln10}. 
Following the procedure adopted in 
\cite{2013MNRAS.432...23G}, in the first $z$-bin all the parameters
have been estimated, whereas, starting from the second bin, the value of
$\alpha$ has been fixed and only $M_{d,\star}$ and $\Phi_d^{\star}$ have
been let free to vary. Table \ref{table1} reports
 the best-fitted parameters, their associated errors and the number of
 galaxies in each bin and in Fig. \ref{figure_vmax} the best-fitted Schechter functions have been
over-plotted to the data points. Concerning the faint-end slope
$\alpha$, our fit yields $\alpha\sim$1.48,
suggesting a larger number of galaxies with small $M_{dust}$ then that obtained by, e.g., \cite{2011MNRAS.417.1510D}.

Since, as previously stated, so far the faint end of the DMF has been poorly investigated 
in our redshift bins, 
we have compared our value for the DMF slope with the ones found in the Local Universe and reported in
\cite{2018MNRAS.479.1077B} (see their Table 1). Faint-end slope values consistent with ours, i. e. between $\alpha\sim 1.3$ and $\alpha\sim 1.4$, 
have been found in the majority of the other data-sets (i.e. \citealt{2005MNRAS.364.1253V},
\citealt{2013MNRAS.433..695C}, \citealt{2018MNRAS.479.1077B}, see
Fig. \ref{figure_dmf_local}), with the exception of 
the work from \cite{2011MNRAS.417.1510D} that finds, as anticipated, a flat faint-end DMF ($\alpha=1.01^{+0.17}_{-0.14}$). 

We find a clear positive 
evolution of $M_{d}^{\star}$ with redshift, indicating that galaxies become more and
more dusty at higher redshift (at $z\sim2$, $M_{d}^{\star}$ is a factor of ten
higher than at $z\sim0.2$). On the contrary, their number
density, expressed in terms of $\Phi_d^{\star}$, remains almost constant up
to $z\sim$0.8, and then drops at higher
redshift. Since at $z>1$ we do not sample the break of
the DMF, at high-$z$ our conclusions need further investigations.
The global evolutionary trends found for $M_{d}^{\star}$ and  $\Phi_d^{\star}$ are in
agreement with the results obtained for the IR luminosity functions (i.e. \citealt{2013MNRAS.432...23G}).
This is somewhat expected given that the two quantities $M_{d}$ and $L_{IR}$
are both related to the galactic star formation activity. 
However, $M_{d}$ and $L_{IR}$ are not simply proportional to each other, since their dependence
on the ISM temperature is different: at a fixed sub-mm flux, $L_{IR}$ increases as the
temperature of the ISM increases, while $M_{d}$ decreases (see also
Appendix C. in \citealp{2014A&A...562A..30S}).  This might 
explain why we do not need to use a modified-Schechter function
(\citealt{1990MNRAS.242..318S}) to reproduce the DMF, as generally
done for the IR luminosity function. In the latter case a simple Schechter parameterization
does not provide a satisfactory fit to the data which remains higher
than the expected exponential decrease of the Schechter function.

The bright end of the far-IR luminosity function is dominated by ULIRGs, characterized by
average high ISM  
temperatures (caused by intense SF activity, perhaps triggered by mergers or by the presence of an AGN).
Such high temperatures certainly have a strong influence on their overall 
IR luminosity, but the bulk of their dust mass is expected to be cold.

\subsection{Dust mass density}
\label{mdust_density_fig}

In Fig. \ref{mdust_density_fig} we report our recovered dust mass density
$\rho_d$ (cyan region) as a function of the lookback time, estimated by integrating the best-fit DMFs
in each of the 6 redshift bins of Fig.4. In each bin the Schechter function has been 
integrated beyond the range over which it has been directly
measured, i.e. down to $M_{d}=10^{4}$ M$_{\odot}$. 
This extrapolation 
depends on the faint-end slope $\alpha$, which was estimated 
only in the first redshift bin ($\alpha\sim1.48$, see 
Sec. \ref{dmf_fit_sec}). To have a feeling of the effect of a different
faint-end slope, we show in the figure also the values we obtain (open blues squares) by fixing $\alpha$ to 1.2
(and re-evaluating $M_{d,\star}$ and $\Phi^{d,\star}$ at each redshift as described in Sect. \ref{dmf_fit_sec}).
The difference is on average a factor $\sim$1.5-2.

In Fig. \ref{mdust_density_fig} we also show previous determinations
for $\rho_d$. The only other results that span a wide redshift range and
take into account the far-IR emission of the dust are those of 
\cite{2018MNRAS.475.2891D}, while the other determinations rely on
indirect measures (e. g., based on stellar masses, or derived from the
optical emission). In comparison to us, \cite{2018MNRAS.475.2891D} the catalogues 
are not based on a blind far-IR selection, but in their
work 
the far-IR fluxes are measured at the position of 
sources detected at lower wavelengths (see \citealt{2017MNRAS.470.1342A} for
details). This leads to a larger catalogue of sources with far-IR
fluxes (i.e. in G10-COSMOS $\sim$24k galaxies in an area of $\sim$1 deg$^2$).
Moreover, \cite{2018MNRAS.475.2891D} correct for
volume-limited effects using a different method than us, i.e. by
fitting a spline to the data before the turn-down due to
incompleteness effects, and integrating under the extrapolated spline
to get a total mass. Despite the different methods used by us and by
\cite{2018MNRAS.475.2891D}, the two analyses show a trend of the
DMD as a function of the lookback time in qualitative agreement. In fact, 
similarly to \cite{2018MNRAS.475.2891D}, we find that the DMD peaks
around 8 Gyr ago (lookback time t$_{lookback}$, corresponding to $z\sim$~1), with an increase between
t$_{lookback}{\sim}$~10 Gyr ($z\sim$~2) and t$_{lookback}{\sim}$~8 Gyr ($z\sim$~1) and 
then smoothly declines from t$_{lookback}{\sim}$~8 Gyr up to the Local Universe. 
Although at t$_{lookback}>6$ Gyr our results are reasonably consistent with \cite{2018MNRAS.475.2891D}, 
at lower redshifts we find that the two estimates differ at 
${\sim}1.5{\sigma}$ level and our findings indicate a shallower
decreasing towards the Local Universe. 

The only other work entirely based on far-IR photometry is the one of 
\cite{2011MNRAS.417.1510D}, spanning a smaller time range
t$_{lookback}<$~4.5 Gyr ($z<$~0.4). Overall, we find that all values are within 2$\sigma$.
Moving to DMD estimates based on indirect observables, our findings are in
agreement with the \cite{2012ApJ...754..116M} estimates from Mg II
absorbers in the spectra of distant quasars (all values are within
1$\sigma$), whereas there are significant differences at
t$_{lookback}>8$ Gyr ($z>1$) with
the estimates obtained by \cite{2018MNRAS.475.2891D} by 
rescaling their stellar mass density with the universal value for dust-to-stellar mass
ratio reported by \cite{2014A&A...567A.103B} (magenta squares in Fig. \ref{mdust_density_fig}).
In general, the dust-to-stellar mass ratio
is expected to vary in galaxies with different star formation
histories (see \citealt{2017MNRAS.465...54C}),
hence the rescaled estimates of  \cite{2018MNRAS.475.2891D} should be regarded with caution. 

As discussed in \cite{2018MNRAS.475.2891D}, a peak in the comoving dust mass density 
at t$_{lookback}{\sim}$~8 Gyr ($z\sim$~1) implies that the build-up of cosmic dust increases in lockstep with the cosmic star formation, and is consistent 
with the observed evolution of the attenuation in galaxies as found in studies of the UV luminosity density between $z=0$ and $z\sim$~4.5
(\citealt{2012A&A...539A..31C}, \citealt{2013A&A...554A..70B}). 
Clearly, the evolution of the DMD at $z>1$ needs to be investigated further, in particular with improved studies of the faint
end of the dust mass function, poorly known at the present time.  
Without any clear knowledge of such quantity, all the present studies might be underestimating the cosmic dust budget 
at high redshift, that could be dominated by the faintest galaxies 
due to their large abundance, as indicated by optical and UV studies
(\citealt{2016ApJ...832...56A}; \citealt{2016MNRAS.456.3194P}), and currently missing from the present sample. 
A significant step forward in this field might come in the future thanks to the SPICA space mission (\citealt{2018PASA...35...30R}). 
With an estimated mid-far IR sensitivity more than an order of
magnitude better than Spitzer and Herschel (\citealt{2017PASA...34...55G}), 
SPICA will allow us a better characterization of the DMF and hopefully a direct measure of its faint-end slope,
potentially with a deep impact on studies of  the evolution of the dust mass density at $z>1$.

In Fig. \ref{mdust_density_fig} we also show the predictions from two
models: the semi-analytic model from \cite{2017MNRAS.471.3152P} \footnote{The results of the cosmological model used here have been
slightly revised by Popping et al. to improve the estimate of
    DMF at high redshift which, as shown in \cite{2017MNRAS.471.3152P},
tends to underestimate the data at $z\sim 2$.}
  and the prediction from the chemical model of \cite{2017MNRAS.471.4615G}.

The semi-analytic model of \cite{2017MNRAS.471.3152P} predicts a
continuous increase of the dust mass density from $z=5$ to $z=1$, followed
by a flat behaviour between $z\sim1$ and $z=0$. The results of this model predict larger dust mass densities than what found in this work,
but are consistent with most of the observational data  at $z~>~1$. At lower redshifts, the model of \cite{2017MNRAS.471.3152P}  predicts an
excess of dust with respect to the amount detected in resolved
galaxies, i.e. in the samples of \cite{2011MNRAS.417.1510D},
\cite{2018MNRAS.475.2891D}, and in our sample, in particular at
$z<0.7$.  As discussed in \cite{2017MNRAS.471.3152P}, the removal of
dust growth on grains in the ISM (see \citealt{2016MNRAS.463L.112F}) alleviates such a tension with the data.

The cosmic chemical evolution model of \cite{2017MNRAS.471.4615G} is based on
models for galaxies of various morphological types
which account for the dust mass budget in local ellipticals, spirals
and irregulars (see also \citealt{2008A&A...479..669C}).
In that model, the redshift evolution of the galaxy populations has been
computed on the basis of a phenomenogical approach (see \citealt{2015ApJ...803...35P}). 
The predictions of  \cite{2017MNRAS.471.4615G} significantly
overestimate our data at $z>1$, however they are consistent with the decrease in the 
comoving dust density at $z<1$ as found in this study.

\section{Summary}
\label{conclusion_sec}

In the literature, only a handful of studies have been performed
on the redshift evolution of the dust mass function. 
We have used a far-IR (160$\mu$m) {\it Herschel} selected catalogue
to derive the DMF across a wide redshift range, 
from $z\sim$~0.2 up to $z\sim$~2.5. 
The starting point of our study was
the latest released blind catalogue selected in the COSMOS field (DR1, 7047 sources) obtained
within the {\it Herschel}-PEP survey
(\citealt{2011A&A...532A..90L}).
From the parent sample of 7047
galaxies, a subsample of 5546 systems presenting a 
counterpart detected at different wavelengths and with an estimate of
the redshift (either spectroscopic or photometric) has been considered for our analysys. 
For each of these systems we have derived a dust mass $M_d$ from the observed
flux at 250 $\mu$m by means of a modified black-body relation.
The dust temperatures have been computed from the empirical
relation found by \cite{2014A&A...561A..86M} for star-forming galaxies
on the basis of their redshift and their specific star formation rate. 
For each dust mass value, we have defined the quantity
$S_{dust}=\frac{M_{dust}}{4{\pi}D_{L}^{2}}$ and we have corrected our sample for incompletess
using the distribution of the $log(S_{160}/S_{dust})$ ratio in various redshift
bins. 
The estimates of the DMF in six different redshift bins
were performed by means of a non-parametric $1/V_{max}$ method.
The DMF computed in each redshift bin was fitted by means of
a Schechter function, and the evolution of its basic parameters was analysed. 
We have integrated our estimates of the 
DMF to study the redshift evolution of the comoving dust mass density up to $z{\sim}2.5$.
We have compared our results with few extant previous studies from the literature. 
Our results can be summarised as follows. 

\begin{itemize}
\item{In our lowest redshift bin (0.1~$<$~$z$~${\leq}$~0.25), our DMF 
is characterised by a fainter bright end and by a flatter faint end
slope with respect to the one obtained by the Phase Verification
Herschel-ATLAS early release \cite{2011MNRAS.417.1510D} while is consistent with the local determination from
\citealt{2018MNRAS.479.1077B}, obtained from the final Herschel-Atlas
analysis. Faint-end slope values consistent 
with ours ($\alpha=1.48$) have been found
in the majority of the other IR-based studies performed in  the local
Universe (\citealt{2018MNRAS.479.1077B} and references therein).}
\item{We find a positive 
  evolution with redshift of the characteristic dust mass $M_{d,\star}$, which
  indicates that galaxies become more and
more dusty at higher redshift (at $z\sim2$, $M_{d}^{\star}$ is a factor of ten
higher than at $z\sim0.2$). On the contrary, their number
density, expressed in terms of the normalization $\Phi_d^{\star}$, stays approximately constant up
to $z\sim$~0.8, and then drops at higher
redshift. The global evolutionary trends found for $M_{d}^{\star}$ and  $\Phi_d^{\star}$ are in
agreement with the results obtained for the IR luminosity function (i.e. \citealt{2013MNRAS.432...23G}).}
\item{The comoving dust mass density peaks
at t$_{lookback}{\sim}$~8 Gyr ($z\sim$~1), with an increase between
t$_{lookback}{\sim}$~10 Gyr ($z\sim$~2) and t$_{lookback}{\sim}$~8 ($z\sim$~1) and 
then it declines from $z\sim$~1 up to the Local Universe. Although at t$_{lookback}>$6 Gys ($z>0.8$) our results are consistent with the ones of \cite{2018MNRAS.475.2891D}, 
at lower redshifts we find that the two estimates differ at
$\sim$1.5$\sigma$ level with our results showing a shallower decrease
toward the local universe. At t$_{lookback}>$6 Gyr ($z>1$), we underline that the
faint-end slope and the break of the DMF are poorly
sampled by Herschel catalogues and, as a consequence, at these redshifts
our conclusions should be regarded with caution. In the future, a
significant step in this field will come from the
 SPICA IR satellite (\citealt{2018PASA...35...30R}) which will observe
 the Universe with an estimated mid-far IR sensitivity more than an order of
magnitude better than Spitzer and Herschel (\citealt{2017PASA...34...55G}).}

\end{itemize}

\begin{table*}
\caption{The Schechter parameters of the DMF. The parameter $\Phi^{\star}$
  incorportares the factor $ln10$.  } 
\begin{tabular}{cccccc}
\hline\hline
\small Redshift range & $\alpha$ & $M_{d}^{\star}$              & $\Phi^{\star}$  & $\rho_{d}$ \\
       &          & [log$_{10}(M_{\odot})$] &      [$10^{-3}$Mpc$^{-3}$dex$^{-1}$]  &   [$10^{5}$ M$_{\odot}$Mpc$^{-3}$] & N objects\\
\hline
0.1-0.25 &  1.48$\pm$0.15 & 7.58 $\pm$0.08 &  4.9$\pm$1.6     &1.32$\pm$0.17 &555\\
0.25-0.4 &  1.48$^{a}$    & 7.80 $\pm$0.03  & 5.0$\pm$0.7     &2.24$\pm$0.29  &801\\
0.4-0.8 &   1.48$^{a}$     & 8.05$\pm$0.02  &  4.4$\pm$0.7     & 3.62$\pm$0.46 &1512 \\
0.8-1.2 &   1.48$^{a}$     & 8.22$\pm$0.07 &    3.3$\pm$0.3    &3.85$\pm$0.48 &1215\\
1.2-1.8 &   1.48$^{a}$     & 8.58$\pm$0.05  & 0.8$\pm$0.1      &2.22$\pm$0.29 &   804\\
1.8-2.5 &   1.48$^{a}$     & 8.91$\pm$0.09  & 0.2$\pm$0.2      &1.38$\pm$0.17 &435 \\
 \hline\hline
$^{a}$ fixed value
\end{tabular}
\label{table1}
\end{table*}


\section{Acknowledgements}
We are grateful to an anonymous referee for valuable suggestions
that improved the paper. FP gratefully thanks Maud Galametz, Andrea Cimatti and Margherita
Talia for useful
discussions. FP thanks Lara Pantoni for comparing the dust masses. 
FP, FC and CG acknowledge funding from the INAFPRIN-SKA 2017 programme
1.05.01.88.04. 

\newpage
\begin{figure*}
\includegraphics[width=0.7\textwidth,angle=270]{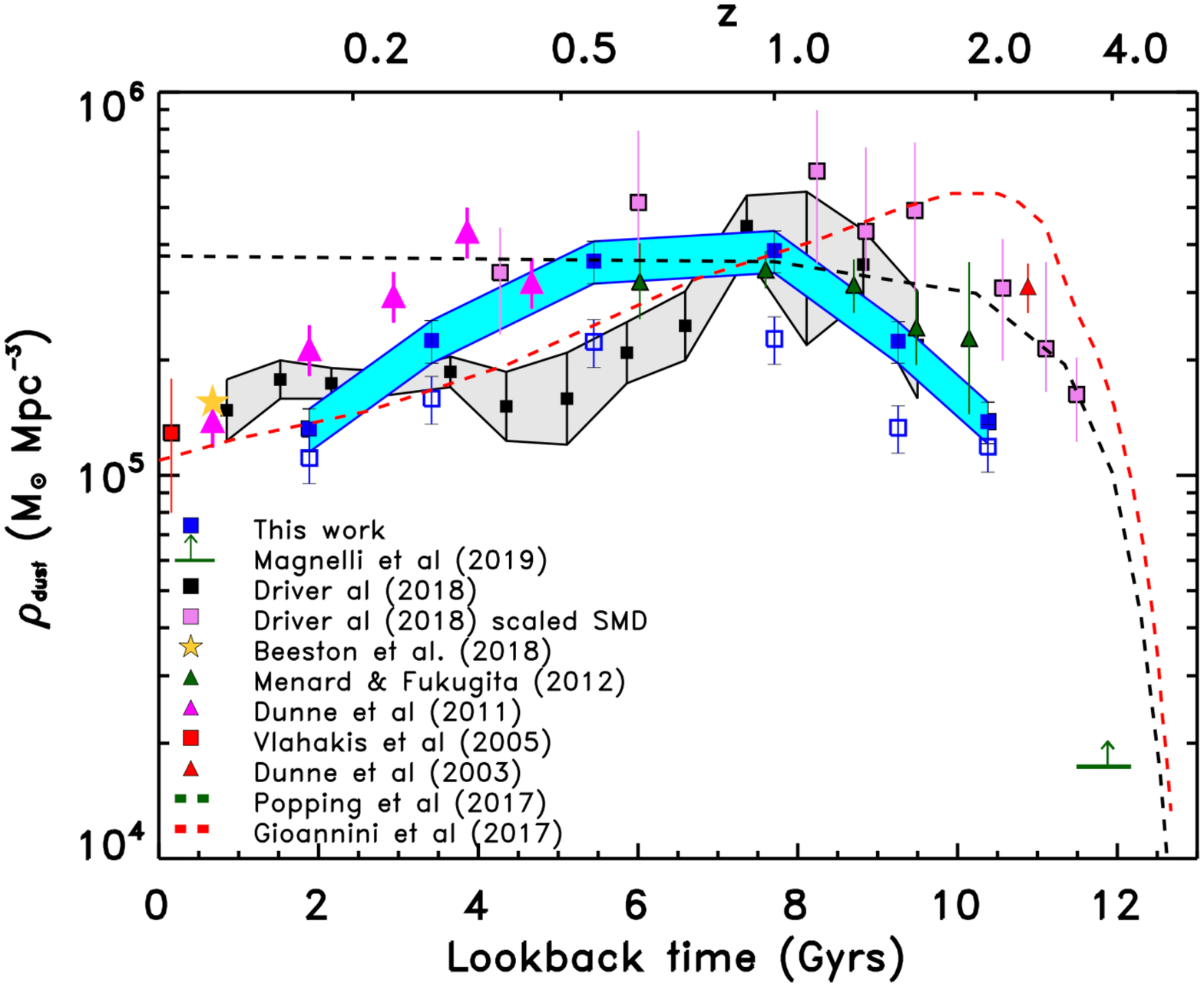}
\caption{DMD versus loookback time estimated by integrating the DMF
  Schechter functions (see parameters in Table 1) in the 6 redshift bins down to 10$^{4}$ M$_{\odot}$
  (Cyan squares). The empty blues squares represent the values
  obtained by fixing the faint-end slope to $\alpha=1.2$. Previous DMD determinations from other authors are also shown, as detailed in the legend. }
\label{mdust_density_fig}
\end{figure*}





\bibliographystyle{mn2e}
\bibliography{pozzi}

\begin{thebibliography}{68}
\expandafter\ifx\csname natexlab\endcsname\relax\def\natexlab#1{#1}\fi

\bibitem[{{Alavi} {et~al}\mbox{.}(2016){Alavi}, {Siana}, {Richard}, {Rafelski},
  {Jauzac}, {Limousin}, {Freeman}, {Scarlata}, {Robertson}, {Stark}, {Teplitz},
  \& {Desai}}]{2016ApJ...832...56A}
{Alavi} A. {et~al.}, 2016, \apj, 832, 56

\bibitem[{{Andrews} {et~al}\mbox{.}(2017){Andrews}, {Driver}, {Davies},
  {Kafle}, {Robotham}, {Vinsen}, {Wright}, {Bland-Hawthorn}, {Bourne},
  {Bremer}, {da Cunha}, {Drinkwater}, {Holwerda}, {Hopkins}, {Kelvin},
  {Loveday}, {Phillipps}, \& {Wilkins}}]{2017MNRAS.470.1342A}
{Andrews} S.~K. {et~al.}, 2017, \mnras, 470, 1342

\bibitem[{{Aoyama} {et~al}\mbox{.}(2018){Aoyama}, {Hou}, {Hirashita},
  {Nagamine}, \& {Shimizu}}]{2018MNRAS.478.4905A}
{Aoyama} S., {Hou} K.-C., {Hirashita} H., {Nagamine} K., {Shimizu} I., 2018,
  \mnras, 478, 4905

\bibitem[{{Arnouts} {et~al}\mbox{.}(2002){Arnouts}, {Moscardini}, {Vanzella},
  {Colombi}, {Cristiani}, {Fontana}, {Giallongo}, {Matarrese}, \&
  {Saracco}}]{2002MNRAS.329..355A}
{Arnouts} S. {et~al.}, 2002, \mnras, 329, 355

\bibitem[{{Beeston} {et~al}\mbox{.}(2018){Beeston}, {Wright}, {Maddox},
  {Gomez}, {Dunne}, {Driver}, {Robotham}, {Clark}, {Vinsen}, {Takeuchi},
  {Popping}, {Bourne}, {Bremer}, {Phillipps}, {Moffett}, {Baes},
  {Bland-Hawthorn}, {Brough}, {De Vis}, {Eales}, {Holwerda}, {Loveday},
  {Liske}, {Smith}, {Smith}, {Valiante}, {Vlahakis}, \&
  {Wang}}]{2018MNRAS.479.1077B}
{Beeston} R.~A. {et~al.}, 2018, \mnras, 479, 1077

\bibitem[{{B{\'e}thermin} {et~al}\mbox{.}(2015){B{\'e}thermin}, {Daddi},
  {Magdis}, {Lagos}, {Sargent}, {Albrecht}, {Aussel}, {Bertoldi}, {Buat},
  {Galametz}, {Heinis}, {Ilbert}, {Karim}, {Koekemoer}, {Lacey}, {Le Floc'h},
  {Navarrete}, {Pannella}, {Schreiber}, {Smol{\v c}i{\'c}}, {Symeonidis}, \&
  {Viero}}]{2015A&A...573A.113B}
{B{\'e}thermin} M. {et~al.}, 2015, \aap, 573, A113

\bibitem[{{B{\'e}thermin} {et~al}\mbox{.}(2014){B{\'e}thermin}, {Kilbinger},
  {Daddi}, {Gabor}, {Finoguenov}, {McCracken}, {Wolk}, {Aussel}, {Strazzulo},
  {Le Floc'h}, {Gobat}, {Rodighiero}, {Dickinson}, {Wang}, {Lutz}, \&
  {Heinis}}]{2014A&A...567A.103B}
{B{\'e}thermin} M. {et~al.}, 2014, \aap, 567, A103

\bibitem[{{Bianchi}(2013)}]{2013A&A...552A..89B}
{Bianchi} S., 2013, \aap, 552, A89

\bibitem[{{Boquien} {et~al}\mbox{.}(2019){Boquien}, {Burgarella}, {Roehlly},
  {Buat}, {Ciesla}, {Corre}, {Inoue}, \& {Salas}}]{2019A&A...622A.103B}
{Boquien} M., {Burgarella} D., {Roehlly} Y., {Buat} V., {Ciesla} L., {Corre}
  D., {Inoue} A.~K., {Salas} H., 2019, \aap, 622, A103

\bibitem[{{Burgarella} {et~al}\mbox{.}(2013){Burgarella}, {Buat}, {Gruppioni},
  {Cucciati}, {Heinis}, {Berta}, {B{\'e}thermin}, {Bock}, {Cooray}, {Dunlop},
  {Farrah}, {Franceschini}, {Le Floc'h}, {Lutz}, {Magnelli}, {Nordon},
  {Oliver}, {Page}, {Popesso}, {Pozzi}, {Riguccini}, {Vaccari}, \&
  {Viero}}]{2013A&A...554A..70B}
{Burgarella} D. {et~al.}, 2013, \aap, 554, A70

\bibitem[{{Burgarella} {et~al}\mbox{.}(2005){Burgarella}, {Buat}, \&
  {Iglesias-P{\'a}ramo}}]{2005MNRAS.360.1413B}
{Burgarella} D., {Buat} V., {Iglesias-P{\'a}ramo} J., 2005, \mnras, 360, 1413

\bibitem[{{Calura} {et~al}\mbox{.}(2008){Calura}, {Pipino}, \&
  {Matteucci}}]{2008A&A...479..669C}
{Calura} F., {Pipino} A., {Matteucci} F., 2008, \aap, 479, 669 (CPM08)

\bibitem[{{Calura} {et~al}\mbox{.}(2017){Calura}, {Pozzi}, {Cresci}, {Santini},
  {Gruppioni}, {Pozzetti}, {Gilli}, {Matteucci}, \&
  {Maiolino}}]{2017MNRAS.465...54C}
{Calura} F. {et~al.}, 2017, \mnras, 465, 54

\bibitem[{{Ciliegi} {et~al}\mbox{.}(2001){Ciliegi}, {Gruppioni}, {McMahon}, \&
  {Rowan-Robinson}}]{2001Ap&SS.276..957C}
{Ciliegi} P., {Gruppioni} C., {McMahon} R., {Rowan-Robinson} M., 2001, \apss,
  276, 957

\bibitem[{{Clark} {et~al}\mbox{.}(2015){Clark}, {Dunne}, {Gomez}, {Maddox}, {De
  Vis}, {Smith}, {Eales}, {Baes}, {Bendo}, {Bourne}, {Driver}, {Dye},
  {Furlanetto}, {Grootes}, {Ivison}, {Schofield}, {Robotham}, {Rowlands},
  {Valiante}, {Vlahakis}, {van der Werf}, {Wright}, \& {de
  Zotti}}]{2015MNRAS.452..397C}
{Clark} C.~J.~R. {et~al.}, 2015, \mnras, 452, 397

\bibitem[{{Clemens} {et~al}\mbox{.}(2013){Clemens}, {Negrello}, {De Zotti},
  {Gonzalez-Nuevo}, {Bonavera}, {Cosco}, {Guarese}, {Boaretto}, {Salucci},
  {Baccigalupi}, {Clements}, {Danese}, {Lapi}, {Mandolesi}, {Partridge},
  {Perrotta}, {Serjeant}, {Scott}, \& {Toffolatti}}]{2013MNRAS.433..695C}
{Clemens} M.~S. {et~al.}, 2013, \mnras, 433, 695

\bibitem[{{Cowie} {et~al}\mbox{.}(1996){Cowie}, {Songaila}, {Hu}, \&
  {Cohen}}]{1996AJ....112..839C}
{Cowie} L.~L., {Songaila} A., {Hu} E.~M., {Cohen} J.~G., 1996, \aj, 112, 839

\bibitem[{{Cucciati} {et~al}\mbox{.}(2012){Cucciati}, {Tresse}, {Ilbert}, {Le
  F{\`e}vre}, {Garilli}, {Le Brun}, {Cassata}, {Franzetti}, {Maccagni},
  {Scodeggio}, {Zucca}, {Zamorani}, {Bardelli}, {Bolzonella}, {Bielby},
  {McCracken}, {Zanichelli}, \& {Vergani}}]{2012A&A...539A..31C}
{Cucciati} O. {et~al.}, 2012, \aap, 539, A31

\bibitem[{{da Cunha} {et~al}\mbox{.}(2008){da Cunha}, {Charlot}, \&
  {Elbaz}}]{2008MNRAS.388.1595D}
{da Cunha} E., {Charlot} S., {Elbaz} D., 2008, \mnras, 388, 1595

\bibitem[{{Davies} {et~al}\mbox{.}(2015){Davies}, {Robotham}, {Driver},
  {Alpaslan}, {Baldry}, {Bland-Hawthorn}, {Brough}, {Brown}, {Cluver},
  {Drinkwater}, {Foster}, {Grootes}, {Konstantopoulos}, {Lara-L{\'o}pez},
  {L{\'o}pez-S{\'a}nchez}, {Loveday}, {Meyer}, {Moffett}, {Norberg}, {Owers},
  {Popescu}, {De Propris}, {Sharp}, {Tuffs}, {Wang}, {Wilkins}, {Dunne},
  {Bourne}, \& {Smith}}]{2015MNRAS.452..616D}
{Davies} L.~J.~M. {et~al.}, 2015, \mnras, 452, 616

\bibitem[{{De Bernardis} \& {Cooray}(2012)}]{2012ApJ...760...14D}
{De Bernardis} F., {Cooray} A., 2012, \apj, 760, 14

\bibitem[{{Delvecchio} {et~al}\mbox{.}(2014){Delvecchio}, {Gruppioni}, {Pozzi},
  {Berta}, {Zamorani}, {Cimatti}, {Lutz}, {Scott}, {Vignali}, {Cresci},
  {Feltre}, {Cooray}, {Vaccari}, {Fritz}, {Le Floc'h}, {Magnelli}, {Popesso},
  {Oliver}, {Bock}, {Carollo}, {Contini}, {Le F{\'e}vre}, {Lilly}, {Mainieri},
  {Renzini}, \& {Scodeggio}}]{2014MNRAS.439.2736D}
{Delvecchio} I. {et~al.}, 2014, \mnras, 439, 2736

\bibitem[{{Delvecchio} {et~al}\mbox{.}(2015){Delvecchio}, {Lutz}, {Berta},
  {Rosario}, {Zamorani}, {Pozzi}, {Gruppioni}, {Vignali}, {Brusa}, {Cimatti},
  {Clements}, {Cooray}, {Farrah}, {Lanzuisi}, {Oliver}, {Rodighiero},
  {Santini}, \& {Symeonidis}}]{2015MNRAS.449..373D}
{Delvecchio} I. {et~al.}, 2015, \mnras, 449, 373

\bibitem[{{Draine}(1990)}]{1990ASPC...12..193D}
{Draine} B.~T., 1990, in Astronomical Society of the Pacific Conference Series,
  Vol.~12, The Evolution of the Interstellar Medium, {Blitz} L., ed., pp.
  193--205

\bibitem[{{Draine}(2009)}]{2009EAS....35..245D}
{Draine} B.~T., 2009, in EAS Publications Series, Vol.~35, EAS Publications
  Series, {Boulanger} F., {Joblin} C., {Jones} A., {Madden} S., eds., pp.
  245--268

\bibitem[{{Draine} \& {Li}(2007)}]{2007ApJ...657..810D}
{Draine} B.~T., {Li} A., 2007, \apj, 657, 810

\bibitem[{{Driver} {et~al}\mbox{.}(2018){Driver}, {Andrews}, {da Cunha},
  {Davies}, {Lagos}, {Robotham}, {Vinsen}, {Wright}, {Alpaslan},
  {Bland-Hawthorn}, {Bourne}, {Brough}, {Bremer}, {Cluver}, {Colless},
  {Conselice}, {Dunne}, {Eales}, {Gomez}, {Holwerda}, {Hopkins}, {Kafle},
  {Kelvin}, {Loveday}, {Liske}, {Maddox}, {Phillipps}, {Pimbblet}, {Rowlands},
  {Sansom}, {Taylor}, {Wang}, \& {Wilkins}}]{2018MNRAS.475.2891D}
{Driver} S.~P. {et~al.}, 2018, \mnras, 475, 2891

\bibitem[{{Driver} {et~al}\mbox{.}(2011){Driver}, {Hill}, {Kelvin}, {Robotham},
  {Liske}, {Norberg}, {Baldry}, {Bamford}, {Hopkins}, {Loveday}, {Peacock},
  {Andrae}, {Bland-Hawthorn}, {Brough}, {Brown}, {Cameron}, {Ching}, {Colless},
  {Conselice}, {Croom}, {Cross}, {de Propris}, {Dye}, {Drinkwater}, {Ellis},
  {Graham}, {Grootes}, {Gunawardhana}, {Jones}, {van Kampen}, {Maraston},
  {Nichol}, {Parkinson}, {Phillipps}, {Pimbblet}, {Popescu}, {Prescott},
  {Roseboom}, {Sadler}, {Sansom}, {Sharp}, {Smith}, {Taylor}, {Thomas},
  {Tuffs}, {Wijesinghe}, {Dunne}, {Frenk}, {Jarvis}, {Madore}, {Meyer},
  {Seibert}, {Staveley-Smith}, {Sutherland}, \& {Warren}}]{2011MNRAS.413..971D}
{Driver} S.~P. {et~al.}, 2011, \mnras, 413, 971

\bibitem[{{Dunne} {et~al}\mbox{.}(2003){Dunne}, {Eales}, \&
  {Edmunds}}]{2003MNRAS.341..589D}
{Dunne} L., {Eales} S.~A., {Edmunds} M.~G., 2003, \mnras, 341, 589

\bibitem[{{Dunne} {et~al}\mbox{.}(2011){Dunne}, {Gomez}, {da Cunha}, {Charlot},
  {Dye}, {Eales}, {Maddox}, {Rowlands}, {Smith}, {Auld}, {Baes}, {Bonfield},
  {Bourne}, {Buttiglione}, {Cava}, {Clements}, {Coppin}, {Cooray}, {Dariush},
  {de Zotti}, {Driver}, {Fritz}, {Geach}, {Hopwood}, {Ibar}, {Ivison},
  {Jarvis}, {Kelvin}, {Pascale}, {Pohlen}, {Popescu}, {Rigby}, {Robotham},
  {Rodighiero}, {Sansom}, {Serjeant}, {Temi}, {Thompson}, {Tuffs}, {van der
  Werf}, \& {Vlahakis}}]{2011MNRAS.417.1510D}
{Dunne} L. {et~al.}, 2011, \mnras, 417, 1510

\bibitem[{{Ferrara} {et~al}\mbox{.}(2016){Ferrara}, {Viti}, \&
  {Ceccarelli}}]{2016MNRAS.463L.112F}
{Ferrara} A., {Viti} S., {Ceccarelli} C., 2016, \mnras, 463, L112

\bibitem[{{Fontana} {et~al}\mbox{.}(2004){Fontana}, {Pozzetti}, {Donnarumma},
  {Renzini}, {Cimatti}, {Zamorani}, {Menci}, {Daddi}, {Giallongo}, {Mignoli},
  {Perna}, {Salimbeni}, {Saracco}, {Broadhurst}, {Cristiani}, {D'Odorico}, \&
  {Gilmozzi}}]{2004A&A...424...23F}
{Fontana} A. {et~al.}, 2004, \aap, 424, 23

\bibitem[{{Fukugita}(2011)}]{2011arXiv1103.4191F}
{Fukugita} M., 2011, arXiv e-prints

\bibitem[{{Fukugita} \& {Peebles}(2004)}]{2004ApJ...616..643F}
{Fukugita} M., {Peebles} P.~J.~E., 2004, \apj, 616, 643

\bibitem[{{Gilli} {et~al}\mbox{.}(2014){Gilli}, {Norman}, {Vignali},
  {Vanzella}, {Calura}, {Pozzi}, {Massardi}, {Mignano}, {Casasola}, {Daddi},
  {Elbaz}, {Dickinson}, {Iwasawa}, {Maiolino}, {Brusa}, {Vito}, {Fritz},
  {Feltre}, {Cresci}, {Mignoli}, {Comastri}, \&
  {Zamorani}}]{2014A&A...562A..67G}
{Gilli} R. {et~al.}, 2014, \aap, 562, A67

\bibitem[{{Gioannini} {et~al}\mbox{.}(2017){Gioannini}, {Matteucci}, \&
  {Calura}}]{2017MNRAS.471.4615G}
{Gioannini} L., {Matteucci} F., {Calura} F., 2017, \mnras, 471, 4615

\bibitem[{{Gruppioni} {et~al}\mbox{.}(2017){Gruppioni}, {Ciesla},
  {Hatziminaoglou}, {Pozzi}, {Rodighiero}, {Santini}, {Armus}, {Baes},
  {Braine}, {Charmandaris}, {Clements}, {Christopher}, {Dannerbauer},
  {Efstathiou}, {Egami}, {Fern{\'a}ndez-Ontiveros}, {Fontanot}, {Franceschini},
  {Gonz{\'a}lez-Alfonso}, {Griffin}, {Kaneda}, {Marchetti}, {Monaco},
  {Nakagawa}, {Onaka}, {Papadopoulos}, {Pearson}, {P{\'e}rez-Fournon},
  {Per{\'e}z-Gonz{\'a}lez}, {Roelfsema}, {Scott}, {Serjeant}, {Spinoglio},
  {Vaccari}, {van der Tak}, {Vignali}, {Wang}, \& {Wada}}]{2017PASA...34...55G}
{Gruppioni} C. {et~al.}, 2017, \pasa, 34, e055

\bibitem[{{Gruppioni} \& {Pozzi}(2019)}]{2019MNRAS.483.1993G}
{Gruppioni} C., {Pozzi} F., 2019, \mnras, 483, 1993

\bibitem[{{Gruppioni} {et~al}\mbox{.}(2013){Gruppioni}, {Pozzi}, {Rodighiero},
  {Delvecchio}, {Berta}, {Pozzetti}, {Zamorani}, {Andreani}, {Cimatti},
  {Ilbert}, {Le Floc'h}, {Lutz}, {Magnelli}, {Marchetti}, {Monaco}, {Nordon},
  {Oliver}, {Popesso}, {Riguccini}, {Roseboom}, {Rosario}, {Sargent},
  {Vaccari}, {Altieri}, {Aussel}, {Bongiovanni}, {Cepa}, {Daddi},
  {Dom{\'{\i}}nguez-S{\'a}nchez}, {Elbaz}, {F{\"o}rster Schreiber}, {Genzel},
  {Iribarrem}, {Magliocchetti}, {Maiolino}, {Poglitsch}, {P{\'e}rez
  Garc{\'{\i}}a}, {Sanchez-Portal}, {Sturm}, {Tacconi}, {Valtchanov},
  {Amblard}, {Arumugam}, {Bethermin}, {Bock}, {Boselli}, {Buat}, {Burgarella},
  {Castro-Rodr{\'{\i}}guez}, {Cava}, {Chanial}, {Clements}, {Conley}, {Cooray},
  {Dowell}, {Dwek}, {Eales}, {Franceschini}, {Glenn}, {Griffin},
  {Hatziminaoglou}, {Ibar}, {Isaak}, {Ivison}, {Lagache}, {Levenson}, {Lu},
  {Madden}, {Maffei}, {Mainetti}, {Nguyen}, {O'Halloran}, {Page}, {Panuzzo},
  {Papageorgiou}, {Pearson}, {P{\'e}rez-Fournon}, {Pohlen}, {Rigopoulou},
  {Rowan-Robinson}, {Schulz}, {Scott}, {Seymour}, {Shupe}, {Smith}, {Stevens},
  {Symeonidis}, {Trichas}, {Tugwell}, {Vigroux}, {Wang}, {Wright}, {Xu},
  {Zemcov}, {Bardelli}, {Carollo}, {Contini}, {Le F{\'e}vre}, {Lilly},
  {Mainieri}, {Renzini}, {Scodeggio}, \& {Zucca}}]{2013MNRAS.432...23G}
{Gruppioni} C. {et~al.}, 2013, \mnras, 432, 23 (GPR13)

\bibitem[{{Hunt} {et~al}\mbox{.}(2019){Hunt}, {Tilunaite}, {Bass}, {Soeller},
  {Llewelyn Roderick}, {Rajagopal}, \& {Crampin}}]{2019arXiv190204851H}
{Hunt} H., {Tilunaite} A., {Bass} G., {Soeller} C., {Llewelyn Roderick} H.,
  {Rajagopal} V., {Crampin} E.~J., 2019, arXiv e-prints

\bibitem[{{Ilbert} {et~al}\mbox{.}(2006){Ilbert}, {Arnouts}, {McCracken},
  {Bolzonella}, {Bertin}, {Le F{\`e}vre}, {Mellier}, {Zamorani}, {Pell{\`o}},
  {Iovino}, {Tresse}, {Le Brun}, {Bottini}, {Garilli}, {Maccagni}, {Picat},
  {Scaramella}, {Scodeggio}, {Vettolani}, {Zanichelli}, {Adami}, {Bardelli},
  {Cappi}, {Charlot}, {Ciliegi}, {Contini}, {Cucciati}, {Foucaud}, {Franzetti},
  {Gavignaud}, {Guzzo}, {Marano}, {Marinoni}, {Mazure}, {Meneux}, {Merighi},
  {Paltani}, {Pollo}, {Pozzetti}, {Radovich}, {Zucca}, {Bondi}, {Bongiorno},
  {Busarello}, {de La Torre}, {Gregorini}, {Lamareille}, {Mathez}, {Merluzzi},
  {Ripepi}, {Rizzo}, \& {Vergani}}]{2006A&A...457..841I}
{Ilbert} O. {et~al.}, 2006, \aap, 457, 841

\bibitem[{{Ilbert} {et~al}\mbox{.}(2009){Ilbert}, {Capak}, {Salvato}, {Aussel},
  {McCracken}, {Sanders}, {Scoville}, {Kartaltepe}, {Arnouts}, {Le Floc'h},
  {Mobasher}, {Taniguchi}, {Lamareille}, {Leauthaud}, {Sasaki}, {Thompson},
  {Zamojski}, {Zamorani}, {Bardelli}, {Bolzonella}, {Bongiorno}, {Brusa},
  {Caputi}, {Carollo}, {Contini}, {Cook}, {Coppa}, {Cucciati}, {de la Torre},
  {de Ravel}, {Franzetti}, {Garilli}, {Hasinger}, {Iovino}, {Kampczyk},
  {Kneib}, {Knobel}, {Kovac}, {Le Borgne}, {Le Brun}, {Le F{\`e}vre}, {Lilly},
  {Looper}, {Maier}, {Mainieri}, {Mellier}, {Mignoli}, {Murayama}, {Pell{\`o}},
  {Peng}, {P{\'e}rez-Montero}, {Renzini}, {Ricciardelli}, {Schiminovich},
  {Scodeggio}, {Shioya}, {Silverman}, {Surace}, {Tanaka}, {Tasca}, {Tresse},
  {Vergani}, \& {Zucca}}]{2009ApJ...690.1236I}
{Ilbert} O. {et~al.}, 2009, \apj, 690, 1236

\bibitem[{{Ilbert} {et~al}\mbox{.}(2013){Ilbert}, {McCracken}, {Le F{\`e}vre},
  {Capak}, {Dunlop}, {Karim}, {Renzini}, {Caputi}, {Boissier}, {Arnouts},
  {Aussel}, {Comparat}, {Guo}, {Hudelot}, {Kartaltepe}, {Kneib}, {Krogager},
  {Le Floc'h}, {Lilly}, {Mellier}, {Milvang-Jensen}, {Moutard}, {Onodera},
  {Richard}, {Salvato}, {Sanders}, {Scoville}, {Silverman}, {Taniguchi},
  {Tasca}, {Thomas}, {Toft}, {Tresse}, {Vergani}, {Wolk}, \&
  {Zirm}}]{2013A&A...556A..55I}
{Ilbert} O. {et~al.}, 2013, \aap, 556, A55

\bibitem[{{Kennicutt} {et~al}\mbox{.}(2011){Kennicutt}, {Calzetti}, {Aniano},
  {Appleton}, {Armus}, {Beir{\~a}o}, {Bolatto}, {Brandl}, {Crocker}, {Croxall},
  {Dale}, {Donovan Meyer}, {Draine}, {Engelbracht}, {Galametz}, {Gordon},
  {Groves}, {Hao}, {Helou}, {Hinz}, {Hunt}, {Johnson}, {Koda}, {Krause},
  {Leroy}, {Li}, {Meidt}, {Montiel}, {Murphy}, {Rahman}, {Rix}, {Roussel},
  {Sandstrom}, {Sauvage}, {Schinnerer}, {Skibba}, {Smith}, {Srinivasan},
  {Vigroux}, {Walter}, {Wilson}, {Wolfire}, \& {Zibetti}}]{2011PASP..123.1347K}
{Kennicutt} R.~C. {et~al.}, 2011, \pasp, 123, 1347

\bibitem[{{Laigle} {et~al}\mbox{.}(2016){Laigle}, {McCracken}, {Ilbert},
  {Hsieh}, {Davidzon}, {Capak}, {Hasinger}, {Silverman}, {Pichon}, {Coupon},
  {Aussel}, {Le Borgne}, {Caputi}, {Cassata}, {Chang}, {Civano}, {Dunlop},
  {Fynbo}, {Kartaltepe}, {Koekemoer}, {Le F{\`e}vre}, {Le Floc'h}, {Leauthaud},
  {Lilly}, {Lin}, {Marchesi}, {Milvang-Jensen}, {Salvato}, {Sanders},
  {Scoville}, {Smolcic}, {Stockmann}, {Taniguchi}, {Tasca}, {Toft}, {Vaccari},
  \& {Zabl}}]{2016ApJS..224...24L}
{Laigle} C. {et~al.}, 2016, \apjs, 224, 24

\bibitem[{{Lutz} {et~al}\mbox{.}(2011){Lutz}, {Poglitsch}, {Altieri},
  {Andreani}, {Aussel}, {Berta}, {Bongiovanni}, {Brisbin}, {Cava}, {Cepa},
  {Cimatti}, {Daddi}, {Dominguez-Sanchez}, {Elbaz}, {F{\"o}rster Schreiber},
  {Genzel}, {Grazian}, {Gruppioni}, {Harwit}, {Le Floc'h}, {Magdis},
  {Magnelli}, {Maiolino}, {Nordon}, {P{\'e}rez Garc{\'{\i}}a}, {Popesso},
  {Pozzi}, {Riguccini}, {Rodighiero}, {Saintonge}, {Sanchez Portal}, {Santini},
  {Shao}, {Sturm}, {Tacconi}, {Valtchanov}, {Wetzstein}, \&
  {Wieprecht}}]{2011A&A...532A..90L}
{Lutz} D. {et~al.}, 2011, \aap, 532, A90

\bibitem[{{Magnelli} {et~al}\mbox{.}(2014){Magnelli}, {Lutz}, {Saintonge},
  {Berta}, {Santini}, {Symeonidis}, {Altieri}, {Andreani}, {Aussel},
  {B{\'e}thermin}, {Bock}, {Bongiovanni}, {Cepa}, {Cimatti}, {Conley}, {Daddi},
  {Elbaz}, {F{\"o}rster Schreiber}, {Genzel}, {Ivison}, {Le Floc'h}, {Magdis},
  {Maiolino}, {Nordon}, {Oliver}, {Page}, {P{\'e}rez Garc{\'{\i}}a},
  {Poglitsch}, {Popesso}, {Pozzi}, {Riguccini}, {Rodighiero}, {Rosario},
  {Roseboom}, {Sanchez-Portal}, {Scott}, {Sturm}, {Tacconi}, {Valtchanov},
  {Wang}, \& {Wuyts}}]{2014A&A...561A..86M}
{Magnelli} B. {et~al.}, 2014, \aap, 561, A86

\bibitem[{{Marchesi} {et~al}\mbox{.}(2016){Marchesi}, {Civano}, {Elvis},
  {Salvato}, {Brusa}, {Comastri}, {Gilli}, {Hasinger}, {Lanzuisi}, {Miyaji},
  {Treister}, {Urry}, {Vignali}, {Zamorani}, {Allevato}, {Cappelluti},
  {Cardamone}, {Finoguenov}, {Griffiths}, {Karim}, {Laigle}, {LaMassa},
  {Jahnke}, {Ranalli}, {Schawinski}, {Schinnerer}, {Silverman}, {Smolcic},
  {Suh}, \& {Trakhtenbrot}}]{2016ApJ...817...34M}
{Marchesi} S. {et~al.}, 2016, \apj, 817, 34

\bibitem[{{Mathis}(1990)}]{1990ARA&A..28...37M}
{Mathis} J.~S., 1990, \araa, 28, 37

\bibitem[{{M{\'e}nard} \& {Fukugita}(2012)}]{2012ApJ...754..116M}
{M{\'e}nard} B., {Fukugita} M., 2012, \apj, 754, 116

\bibitem[{{Momcheva} {et~al}\mbox{.}(2016){Momcheva}, {Brammer}, {van Dokkum},
  {Skelton}, {Whitaker}, {Nelson}, {Fumagalli}, {Maseda}, {Leja}, {Franx},
  {Rix}, {Bezanson}, {Da Cunha}, {Dickey}, {F{\"o}rster Schreiber},
  {Illingworth}, {Kriek}, {Labb{\'e}}, {Ulf Lange}, {Lundgren}, {Magee},
  {Marchesini}, {Oesch}, {Pacifici}, {Patel}, {Price}, {Tal}, {Wake}, {van der
  Wel}, \& {Wuyts}}]{2016ApJS..225...27M}
{Momcheva} I.~G. {et~al.}, 2016, \apjs, 225, 27

\bibitem[{{Mortlock} {et~al}\mbox{.}(2011){Mortlock}, {Conselice}, {Bluck},
  {Bauer}, {Gr{\"u}tzbauch}, {Buitrago}, \& {Ownsworth}}]{2011MNRAS.413.2845M}
{Mortlock} A., {Conselice} C.~J., {Bluck} A.~F.~L., {Bauer} A.~E.,
  {Gr{\"u}tzbauch} R., {Buitrago} F., {Ownsworth} J., 2011, \mnras, 413, 2845

\bibitem[{{Parsa} {et~al}\mbox{.}(2016){Parsa}, {Dunlop}, {McLure}, \&
  {Mortlock}}]{2016MNRAS.456.3194P}
{Parsa} S., {Dunlop} J.~S., {McLure} R.~J., {Mortlock} A., 2016, \mnras, 456,
  3194

\bibitem[{{Popping} {et~al}\mbox{.}(2017){Popping}, {Somerville}, \&
  {Galametz}}]{2017MNRAS.471.3152P}
{Popping} G., {Somerville} R.~S., {Galametz} M., 2017, \mnras, 471, 3152

\bibitem[{{Pozzi} {et~al}\mbox{.}(2015){Pozzi}, {Calura}, {Gruppioni},
  {Granato}, {Cresci}, {Silva}, {Pozzetti}, {Matteucci}, \&
  {Zamorani}}]{2015ApJ...803...35P}
{Pozzi} F. {et~al.}, 2015, \apj, 803, 35

\bibitem[{{Roelfsema} {et~al}\mbox{.}(2018){Roelfsema}, {Shibai}, {Armus},
  {Arrazola}, {Audard}, {Audley}, {Bradford}, {Charles}, {Dieleman}, {Doi},
  {Duband}, {Eggens}, {Evers}, {Funaki}, {Gao}, {Giard}, {di Giorgio},
  {Gonz{\'a}lez Fern{\'a}ndez}, {Griffin}, {Helmich}, {Hijmering}, {Huisman},
  {Ishihara}, {Isobe}, {Jackson}, {Jacobs}, {Jellema}, {Kamp}, {Kaneda},
  {Kawada}, {Kemper}, {Kerschbaum}, {Khosropanah}, {Kohno}, {Kooijman},
  {Krause}, {van der Kuur}, {Kwon}, {Laauwen}, {de Lange}, {Larsson}, {van
  Loon}, {Madden}, {Matsuhara}, {Najarro}, {Nakagawa}, {Naylor}, {Ogawa},
  {Onaka}, {Oyabu}, {Poglitsch}, {Reveret}, {Rodriguez}, {Spinoglio}, {Sakon},
  {Sato}, {Shinozaki}, {Shipman}, {Sugita}, {Suzuki}, {van der Tak}, {Torres
  Redondo}, {Wada}, {Wang}, {Wafelbakker}, {van Weers}, {Withington},
  {Vandenbussche}, {Yamada}, \& {Yamamura}}]{2018PASA...35...30R}
{Roelfsema} P.~R. {et~al.}, 2018, \pasa, 35, e030

\bibitem[{{Roseboom} {et~al}\mbox{.}(2010){Roseboom}, {Oliver}, {Kunz},
  {Altieri}, {Amblard}, {Arumugam}, {Auld}, {Aussel}, {Babbedge},
  {B{\'e}thermin}, {Blain}, {Bock}, {Boselli}, {Brisbin}, {Buat}, {Burgarella},
  {Castro-Rodr{\'{\i}}guez}, {Cava}, {Chanial}, {Chapin}, {Clements}, {Conley},
  {Conversi}, {Cooray}, {Dowell}, {Dwek}, {Dye}, {Eales}, {Elbaz}, {Farrah},
  {Fox}, {Franceschini}, {Gear}, {Glenn}, {Solares}, {Griffin}, {Halpern},
  {Harwit}, {Hatziminaoglou}, {Huang}, {Ibar}, {Isaak}, {Ivison}, {Lagache},
  {Levenson}, {Lu}, {Madden}, {Maffei}, {Mainetti}, {Marchetti}, {Marsden},
  {Mortier}, {Nguyen}, {O'Halloran}, {Omont}, {Page}, {Panuzzo},
  {Papageorgiou}, {Patel}, {Pearson}, {P{\'e}rez-Fournon}, {Pohlen},
  {Rawlings}, {Raymond}, {Rigopoulou}, {Rizzo}, {Rowan-Robinson}, {Portal},
  {Schulz}, {Scott}, {Seymour}, {Shupe}, {Smith}, {Stevens}, {Symeonidis},
  {Trichas}, {Tugwell}, {Vaccari}, {Valtchanov}, {Vieira}, {Vigroux}, {Wang},
  {Ward}, {Wright}, {Xu}, \& {Zemcov}}]{2010MNRAS.409...48R}
{Roseboom} I.~G. {et~al.}, 2010, \mnras, 409, 48

\bibitem[{{Santini} {et~al}\mbox{.}(2014){Santini}, {Maiolino}, {Magnelli},
  {Lutz}, {Lamastra}, {Li Causi}, {Eales}, {Andreani}, {Berta}, {Buat},
  {Cooray}, {Cresci}, {Daddi}, {Farrah}, {Fontana}, {Franceschini}, {Genzel},
  {Granato}, {Grazian}, {Le Floc'h}, {Magdis}, {Magliocchetti}, {Mannucci},
  {Menci}, {Nordon}, {Oliver}, {Popesso}, {Pozzi}, {Riguccini}, {Rodighiero},
  {Rosario}, {Salvato}, {Scott}, {Silva}, {Tacconi}, {Viero}, {Wang}, {Wuyts},
  \& {Xu}}]{2014A&A...562A..30S}
{Santini} P. {et~al.}, 2014, \aap, 562, A30

\bibitem[{{Saunders} {et~al}\mbox{.}(1990){Saunders}, {Rowan-Robinson},
  {Lawrence}, {Efstathiou}, {Kaiser}, {Ellis}, \&
  {Frenk}}]{1990MNRAS.242..318S}
{Saunders} W., {Rowan-Robinson} M., {Lawrence} A., {Efstathiou} G., {Kaiser}
  N., {Ellis} R.~S., {Frenk} C.~S., 1990, \mnras, 242, 318

\bibitem[{{Savage} \& {Sembach}(1996)}]{1996ARA&A..34..279S}
{Savage} B.~D., {Sembach} K.~R., 1996, \araa, 34, 279

\bibitem[{{Schechter}(1976)}]{1976ApJ...203..297S}
{Schechter} P., 1976, \apj, 203, 297

\bibitem[{{Schmidt}(1968)}]{1968ApJ...151..393S}
{Schmidt} M., 1968, \apj, 151, 393

\bibitem[{{Schreiber} {et~al}\mbox{.}(2018){Schreiber}, {Elbaz}, {Pannella},
  {Ciesla}, {Wang}, \& {Franco}}]{2018A&A...609A..30S}
{Schreiber} C., {Elbaz} D., {Pannella} M., {Ciesla} L., {Wang} T., {Franco} M.,
  2018, \aap, 609, A30

\bibitem[{{Silva} {et~al}\mbox{.}(1998){Silva}, {Granato}, {Bressan}, \&
  {Danese}}]{1998ApJ...509..103S}
{Silva} L., {Granato} G.~L., {Bressan} A., {Danese} L., 1998, \apj, 509, 103

\bibitem[{{Speagle} {et~al}\mbox{.}(2014){Speagle}, {Steinhardt}, {Capak}, \&
  {Silverman}}]{2014ApJS..214...15S}
{Speagle} J.~S., {Steinhardt} C.~L., {Capak} P.~L., {Silverman} J.~D., 2014,
  \apjs, 214, 15

\bibitem[{{Sutherland} \& {Saunders}(1992)}]{1992MNRAS.259..413S}
{Sutherland} W., {Saunders} W., 1992, \mnras, 259, 413

\bibitem[{{Tielens} \& {Allamandola}(1987)}]{1987ppic.proc..333T}
{Tielens} A.~G.~G.~M., {Allamandola} L.~J., 1987, in NATO ASIC Proc. 210:
  Physical Processes in Interstellar Clouds, {Morfill} G.~E., {Scholer} M.,
  eds., pp. 333--376

\bibitem[{{Vlahakis} {et~al}\mbox{.}(2005){Vlahakis}, {Dunne}, \&
  {Eales}}]{2005MNRAS.364.1253V}
{Vlahakis} C., {Dunne} L., {Eales} S., 2005, \mnras, 364, 1253

\end{thebibliography}








\bsp	
\label{lastpage}
\end{document}